\documentclass[aps,preprintnumbers,showpacs,nofootinbib,superscriptaddress,floatfix]{revtex4}

\usepackage{graphicx}
\usepackage{amssymb}
\usepackage{amsmath}
\usepackage{bm}
\usepackage{slashed}
\usepackage{hyperref}
\usepackage{datetime}
\usepackage{mciteplus}
\usepackage{multirow}
\usepackage{color}
\usepackage[utf8]{inputenc}
\usepackage[normalem]{ulem}

\newcommand\as{\alpha_{s}} 
\newcommand\f[2]{\frac{#1}{#2}} 
\def\la{\lambda} 
\def\b0{b_0}

\def\beq{\begin{equation}} 
\def\eeq{\end{equation}} 
\def\beeq{\begin{eqnarray}} 
\def\eeeq{\end{eqnarray}} 
 
\def\to{\rightarrow}
 
\def\nn{\nonumber}

\def\pt{q_{T}}
\def\ptsq{q_{T}^2}
\def\mT{m_{T}}
\def\sig{\sigma}

\def\sh{\hat s}

\def\yT{y_{T}}
\def\yTh{\hat y_{T}}

\def\gev2{{\rm GeV}^2}
\def\nn{\nonumber}

\def\rar{\rightarrow}

\def\lapproxeq{{\ \lower 0.6ex \hbox{$\buildrel<\over\sim$}\ }}
\def\gapproxeq{{\ \lower 0.6ex \hbox{$\buildrel>\over\sim$}\ }}

\def \be  {\begin{equation}}
\def \ee  {\end{equation}}
\def \ba  {\begin{eqnarray}}
\def \ea  {\end{eqnarray}}
\def \baa {\begin{eqnarray*}}
\def \eaa {\end{eqnarray*}}
\def \bb  {\begin{thebibliography}}
\def \eb  {\end{thebibliography}}

\begin{document}

\title{Difficulties in the description of 
Drell--Yan processes at moderate invariant mass and high transverse momentum}

\author{Alessandro Bacchetta}
\email{alessandro.bacchetta@unipv.it}
\affiliation{Dipartimento di Fisica, Universit\`a di Pavia, via Bassi 6, I-27100 Pavia} 
\affiliation{INFN Sezione di Pavia, via Bassi 6, I-27100 Pavia, Italy}

\author{Giuseppe Bozzi}
\email{giuseppe.bozzi@unipv.it}
\affiliation{Dipartimento di Fisica, Universit\`a di Pavia, via Bassi 6, I-27100 Pavia} 
\affiliation{INFN Sezione di Pavia, via Bassi 6, I-27100 Pavia, Italy}

\author{Martin Lambertsen}
\email{lambertsen@tphys.physik.uni-tuebingen.de}
\affiliation{Institute for Theoretical Physics, T\"ubingen University, 
Auf der Morgenstelle 14, D-72076 T\"ubingen, Germany}

\author{Fulvio Piacenza}
\email{fulvio.piacenza01@universitadipavia.it}
\affiliation{Dipartimento di Fisica, Universit\`a di Pavia, via Bassi 6, I-27100 Pavia} 
\affiliation{INFN Sezione di Pavia, via Bassi 6, I-27100 Pavia, Italy}

\author{Julius Steiglechner}
\email{steiglju@gmail.com}
\affiliation{Institute for Theoretical Physics, T\"ubingen University, 
Auf der Morgenstelle 14, D-72076 T\"ubingen, Germany}

\author{Werner Vogelsang}
\email{werner.vogelsang@uni-tuebingen.de}
\affiliation{Institute for Theoretical Physics, T\"ubingen University, 
Auf der Morgenstelle 14, D-72076 T\"ubingen, Germany}

\begin{abstract}
We study the Drell--Yan cross section differential with respect to the transverse momentum of the produced lepton pair. We consider data with 
moderate invariant mass $Q$ of the lepton pair, between 4.5 GeV and 13.5 GeV, and similar (although slightly smaller) values of the transverse momentum $q_T$. We approach the problem by deriving predictions based on standard collinear factorization, which are expected to
be valid toward the high-$q_T$ end of the spectrum and to which any description of the spectrum at lower $q_T$ 
using transverse-momentum dependent parton distributions ultimately needs to be matched. We find that the collinear framework
predicts cross sections that in most cases are significantly below available data at high $q_T$. We discuss additional perturbative and 
possible non-perturbative effects that increase the predicted cross section, but not by a sufficient amount.
\end{abstract}

\preprint{INT-PUB-19-002}
\maketitle

\section{Introduction}
\label{intro}

The Drell--Yan (DY) process~\cite{Drell:1970wh} is one of the main sources of information about the internal structure of the nucleon (for a recent review, see \cite{Peng:2014hta}). Factorization theorems were first established for DY~\cite{Collins:1989gx}, and global extractions of parton distribution functions (PDFs) heavily rely on measurements of the DY cross section differential in the rapidity of the produced boson (see, e.g., \cite{Butterworth:2015oua,Accardi:2016ndt} and references therein). DY processes also offer the possibility to access transverse momentum distributions (TMDs)~\cite{Landry:1999an,Landry:2002ix,DAlesio:2004eso,Konychev:2005iy,DAlesio:2014mrz,Su:2014wpa,Pasquini:2014ppa,Bacchetta:2017gcc,Scimemi:2017etj,Ceccopieri:2018nop}, if the cross section is kept differential in the transverse momentum of the produced boson. 

Considering the invariant mass of the produced boson, $Q$, its transverse momentum, $q_T$, and a typical QCD scale, $\Lambda_{\rm QCD}$, we can distinguish a region of ``high transverse momentum''\footnote{Note that sometimes also $q_T\gg Q$ is 
referred to as the Drell--Yan high transverse momentum regime; see Ref.~\cite{Berger:2001wr}. This regime is usually not accessible
in fixed-target scattering and will therefore not be addressed in the present paper.}  
where $\Lambda_{\rm  QCD} \ll q_T \sim Q$ and a 
region of ``low transverse momentum'' where $q_T \ll Q$. In the first region, the cross section should be well described by a collinear factorization framework in terms of collinear PDFs convoluted with a partonic hard scattering calculated up to a fixed order in $\alpha_s$. This calculation is nowadays possible even up to order $\alpha_s^3$ (NNLO)~\cite{Ridder:2016nkl}, but most of the phenomenology is carried out at order $\alpha_s^2$ (NLO)~\cite{Gonsalves:1989ar,Mirkes:1992hu,Melnikov:2006kv,Catani:2009sm,Gavin:2010az,Bozzi:2010xn,Bozzi:2008bb,Becher:2011xn} or even only order $\alpha_s$ (LO). 

In the low transverse momentum region, the cross section should be described in the framework of TMD factorization, which also incorporates the effects of the resummation of large logarithms in $q_T/Q$. The all-order corrections dominating the cross section in this region are embodied in the 
so-called ``$W$ term'' of the Collins-Soper-Sterman formalism \cite{Collins:1984kg}. The matching of the collinear formalism at high-$q_{T}$ with the TMD resummation at low-$q_{T}$ is usually performed through the introduction of the so-called ``$Y$ term'', i.e., the difference of the fixed-order perturbative result and the asymptotic expansion of the resummed result. In the low-$q_{T}$ region, the asymptotic piece and the fixed-order one ideally cancel each other, leaving only the $W$ term. In the high-$q_{T}$ region, on the other hand, the cancellation takes place between the asymptotic piece and the $W$ term. The situation is more complicated if also the angular dependence of the DY cross section is taken into consideration (see, e.g., \cite{Boer:2006eq,Berger:2007si,Peng:2015spa,Lambertsen:2016wgj}). 

Both regimes, $q_T \ll Q$ and $q_T \sim Q$, as well as their matching, must be under theoretical control in order to have a proper
understanding of the physics of the Drell--Yan process. In the present work, we study the process at fixed-target energies for 
moderate values of the invariant mass $Q$ and in the region $q_T \lesssim Q$. We focus on the predictions based on collinear factorization
and examine their ability to describe the experimental data in this regime. We find in fact that the predicted 
cross sections fall significantly short of the available data even at the highest accessible values of $q_T$. We investigate 
possible sources of uncertainty in the predictions based on collinear factorization, and two extensions of the collinear 
framework: the resummation of high-$q_T$ threshold logarithms, and transverse-momentum smearing. None of these
appears to lead to a satisfactory agreement with the data. We argue that these findings also imply that the Drell--Yan 
cross section in the ``matching regime'' $q_T\lesssim Q$ is presently not fully understood at fixed-target energies.

We note that a similar problem has been reported in~\cite{Gonzalez-Hernandez:2018ipj} for semi-inclusive deep inelastic scattering (SIDIS) processes in the region of large transverse momenta, where large disagreements have been observed also in this case between fixed-order calculations and experimental data. The discrepancies we report here arguably appear more serious since the calculation of the Drell--Yan cross section relies on the very well constrained PDFs, while SIDIS is also sensitive to the comparably more poorly known fragmentation functions.

\section{Motivation: from TMDs to the matching regime}

As described in the Introduction, the regime $q_T \ll Q$  may be addressed in terms
of TMD factorization, and numerous studies using fixed-target Drell--Yan data
have been carried out~\cite{Landry:1999an,Landry:2002ix,DAlesio:2004eso,Konychev:2005iy,DAlesio:2014mrz,Su:2014wpa,Pasquini:2014ppa,Bacchetta:2017gcc,Scimemi:2017etj,Ceccopieri:2018nop}, which however only address the region $q_T\lesssim 1.5$~GeV
and do not make any attempt to perform a matching to a fixed-order calculation at higher $q_T$.  Indeed, extending the description to the whole $q_T$-spectrum is a delicate task. To understand the related issues, it is worth to summarize here the basic ideas behind the most common matching procedures (for detailed expositions, we refer the reader to dedicated studies, e.g., \cite{Collins:1984kg,Arnold:1990yk}). 

The low-$q_T$ formula for the cross section, which embodies TMD physics, has the following expression: 
\begin{equation}
\label{Wterm}
\frac{d\sigma}{dq_T} \propto W \left(q_T \right) = \int \frac{d^2 \bold{b}}{\left(2\pi\right)^2}\,
{\mathrm e}^{i\bold{b \cdot q _T}}W_{\mathrm{pert}} \left(x_A, x_B, b^* \left(b\right), Q \right) W_{{\mathrm{NP}}}\left(x_A, x_B, b, Q \right),
\end{equation} 
where $W_{{\mathrm{pert}}}$ contains soft gluon resummation and $W_{{\mathrm{NP}}}$ the non-perturbative terms. The observed transverse momentum distribution is thus a convolution of the two contributions. Since perturbative calculations would hit the Landau pole at large values of $b$, one common solution is to freeze the impact parameter $b$ beyond a threshold $b_{{\mathrm{max}}}$, by introducing the function $b^*\left(b\right)$, constructed in such a way that $b^*\simeq b$ when $b \ll b_{{\mathrm{max}}}$, and $b^* = b_{{\mathrm{max}}}$ when $b > b_{{\mathrm{max}}}$.

With increasing $q_T$, one expects a smooth transition from TMD physics to collinear factorization. A common way to describe this transition is the following: a correction term (so called ``Y-term") is added to Eq.~(\ref{Wterm}), in order to approximate the sub-leading (in powers of $q_T/Q$) contributions that are not present in the resummed formula.  It is given by the difference between the fixed-order and asymptotic cross sections: 
\begin{equation}
\label{Yterm}
Y(q_T) \propto \frac{d\sigma }{dq_T}^{{\mathrm{(f.o.)}}}- \, \frac{d\sigma }{dq_T}^{{\mathrm{(asy)}}},
\end{equation}
where  the asymptotic piece is obtained by isolating the terms in the fixed-order expression that are most divergent for $q_T/Q \rightarrow 0$. In an ideal situation, at some point as $q_T$ increases towards $Q$, the asymptotic term in Eq.~(\ref{Yterm}) cancels with the $W$ term, so that the sum $W + Y$ approaches the fixed-order cross section  (see, e.g., Sec.~1.4 of \citep{Arnold:1990yk}).

\begin{figure}
\begin{centering}
\includegraphics[scale=0.45]{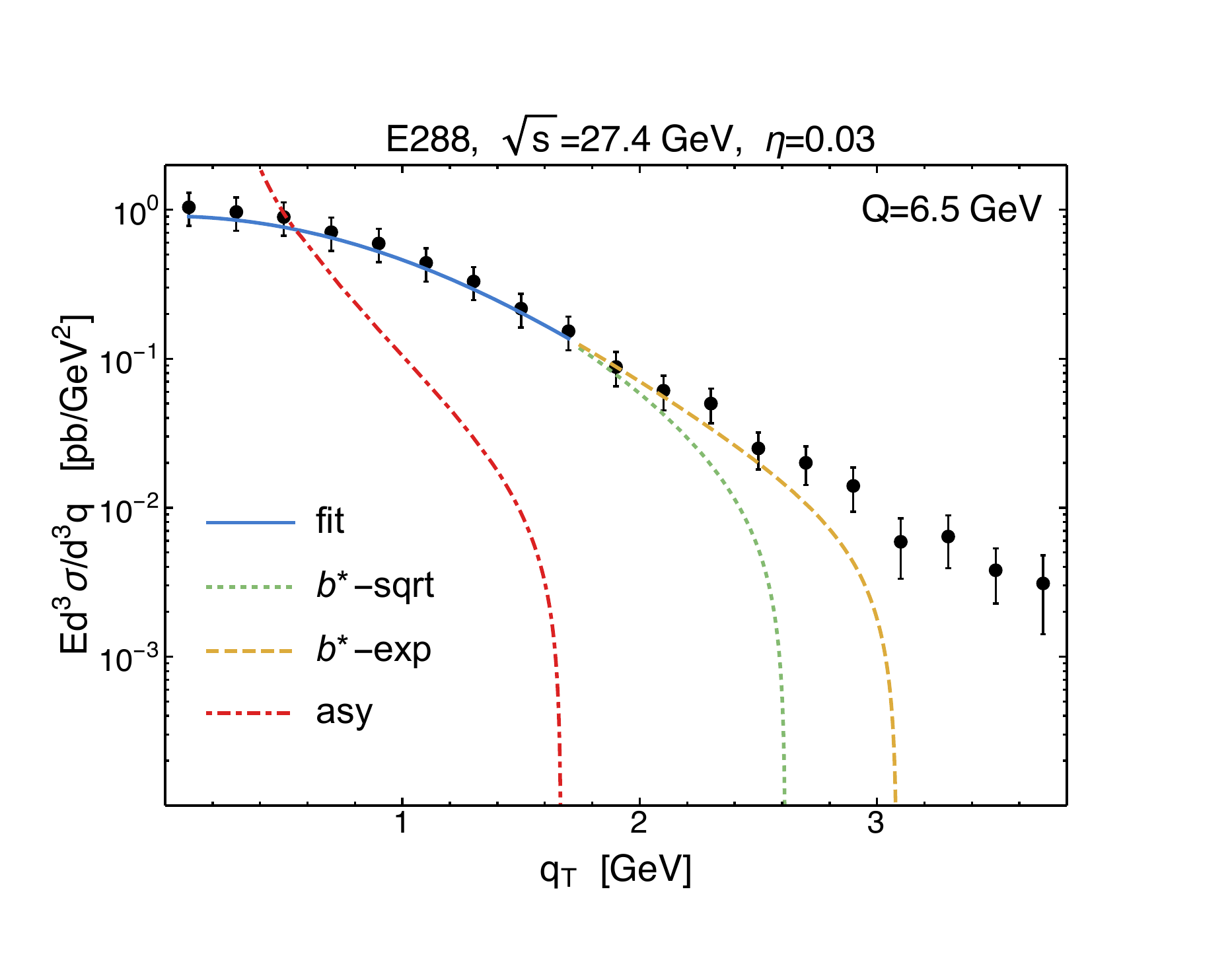}
\includegraphics[scale=0.45]{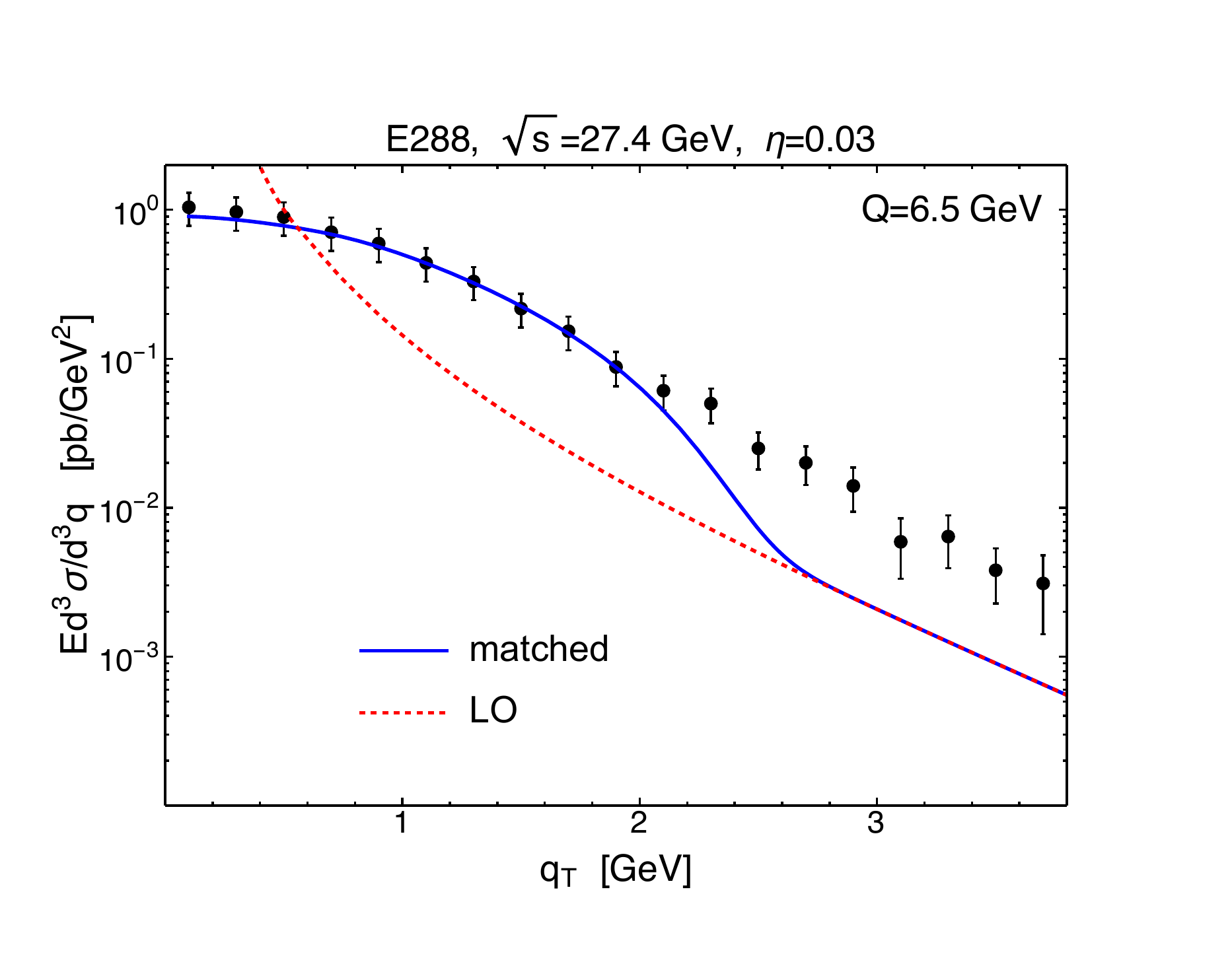}
\end{centering}
\caption{\label{TMDfig} \textbf{Left}: the TMD cross section (full line) from the fit in \cite{Bacchetta:2017gcc}, when extended beyond the fit region, shows markedly different behavior depending on the functional form chosen for $b^*$ in Eq.~(\ref{Wterm}): the dotted line is obtained with a square-root form, while the dashed line with an exponential form (respectively Eqs. 3.18 and 3.19 of \cite{Bacchetta:2015ora}). $b_{{\mathrm{max}}}$ is kept fixed at 1.123 GeV$^{-1}$. The asymptotic curve is also plotted (at LO, to be consistent with the fit). \textbf{Right}: matched curve obtained from the same TMD, with the procedure described in \cite{Collins:2016hqq}. The damping functions are taken as in Sec. IX of the same article. Data are taken from \cite{Ito:1980ev}.}
\end{figure}

The matching procedure can pose serious problems when  $Q$ is not very high, as was shown in \cite{Boglione:2014oea} for the case of SIDIS.\footnote{It is striking that the problems were found to persist even to HERA-like kinematics, with $Q^2 = 100$ GeV$^2$ and $\sqrt{s} = 300$ GeV.}
The observed problems can be summarized as follows: the high-$q_T$ tail of the TMD formula shows sensitivity to the non-perturbative parameters and to the details of the $b^*$ function, preventing a proper cancellation with the $Y$ term. It is straightforward to check that this behavior is 
also present in Drell--Yan at fixed-target kinematics. To give an example, in Fig.~\ref{TMDfig} we show the effect of extrapolating the TMD fitted in \cite{Bacchetta:2017gcc} to high $q_T$. Although the asymptotic curve drops very rapidly at some point, signaling that $\mathcal{O}\left( q_T/Q \right)$ corrections should become dominant, the TMD extends far beyond, owing to the non-perturbative Sudakov contribution.  
We note that also the behavior of the $W_{{\mathrm{pert}}}$ and $W_{{\mathrm{NP}}}$ for $b\to 0$ is expected to 
play a role here. All in all, while on the one side the shape of the data seems to suggest that TMD physics is indeed involved in some form up
to transverse momenta as high as $2.5$~GeV, one has to admit at the same time that presently there is not a good understanding of the TMD
formalism in this region. The matching procedure is afflicted by large uncertainties, and the TMD tail is largely affected by nonperturbative elements, such as the functional form of $b^*$ (see figure for details).

To avoid these (and other) problems, the authors in \cite{Collins:2016hqq}
proposed a modified matching procedure. Without entering into details, we only
underline that this procedure forces the use of pure fixed-order calculation
at intermediate values of $q_T$, by suppressing  the tail of the TMD cross
section with a damping function.  Just to give a qualitative example, in
Fig.~\ref{TMDfig} we show the effect of using the same damping function in our
case. An alternative approach to suppress the TMD contribution at high
transverse momentum was proposed also in \cite{Echevarria:2018qyi}.

In conclusion, the Drell--Yan $q_T$ spectra at low invariant mass are presently not understood beyond the region $q_T \ll Q$ typical of TMD fits. In the following, we will approach the problem from high  $q_T \sim Q$, where collinear factorization is expected to offer a suitable framework for describing the cross section. Undoubtedly the collinear-factorized cross section will be an important ingredient for a better understanding of the regime $q_T\lesssim Q$, where it will be especially important for carrying out the proper matching of the resummed cross section.

\section{Collinear factorization and comparison to fixed-target data \label{coll}}
\label{numerical}

In this section we show the comparison of fixed-order perturbative QCD calculations to Drell--Yan data from Fermilab, CERN and RHIC experiments, mainly for proton-proton collisions. The center-of-mass energies of the experiments taken into account lie in the range 20 GeV $\leq\sqrt{s}\leq$ 60 GeV (except for RHIC, where $\sqrt{s}$=200 GeV), while the invariant mass of the Drell--Yan lepton pair lies in the range $4.5 \leq Q\leq13.5$ GeV.  For all our theoretical predictions, we use the DYqT \cite{Bozzi:2010xn,Bozzi:2008bb} and CuTe \cite{Becher:2011xn} codes, obtaining completely equivalent results for the fixed-order differential cross sections, at both LO QCD $\left(\mathcal{O}\left(\alpha_{s}\right)\right)$ and NLO QCD 
$\left(\mathcal{O}\left(\alpha_{s}^{2}\right)\right)$. 
These codes also provide an all-order resummation of logarithms in $q_T/Q$ in the cross section, which become relevant toward low $q_T$. This enables us to study the asymptotic expansion of the resummed result, which we will make use of below. 
We note that we have also performed cross-checks using the numerical codes of Refs.~\cite{Catani:2009sm} and~\cite{Gavin:2010az}. Throughout this paper, the CT14 PDF set \cite{Dulat:2015mca} will be our default choice. 

\subsection*{E866}

The E866/NuSea experiment \cite{Hawker:1998ty} was a fixed-target Drell--Yan experiment designed to measure the internal structure of the nucleon, in particular the asymmetry of down and up antiquarks in the sea, using di-muon events originating from the collision of an 800-GeV proton beam with hydrogen and deuterium targets ($\sqrt s$ = 38.8 GeV). The measurement of the $q_T$-distribution of the muon pair is presented in \cite{Webb:2003bj}, a Fermilab PhD thesis, and results are given in terms of the differential cross section:
\begin{equation} 
\frac{Ed^{3}\sigma}{d^{3}q}\equiv\frac{2E}{\pi\sqrt{s}}\frac{d\sigma}{dx_{F}dq_T^{2}}=\frac{d\sigma}{\pi dydq_T^{2}}\,.
\end{equation} 
Data are reported for different bins in $x_{F} = 2p_L/\sqrt s$, ranging from $-0.05$ to 0.8, and are integrated over different ranges in the invariant mass $Q$ of the muon pair.

\begin{figure}
\begin{centering}
\includegraphics[scale=0.45]{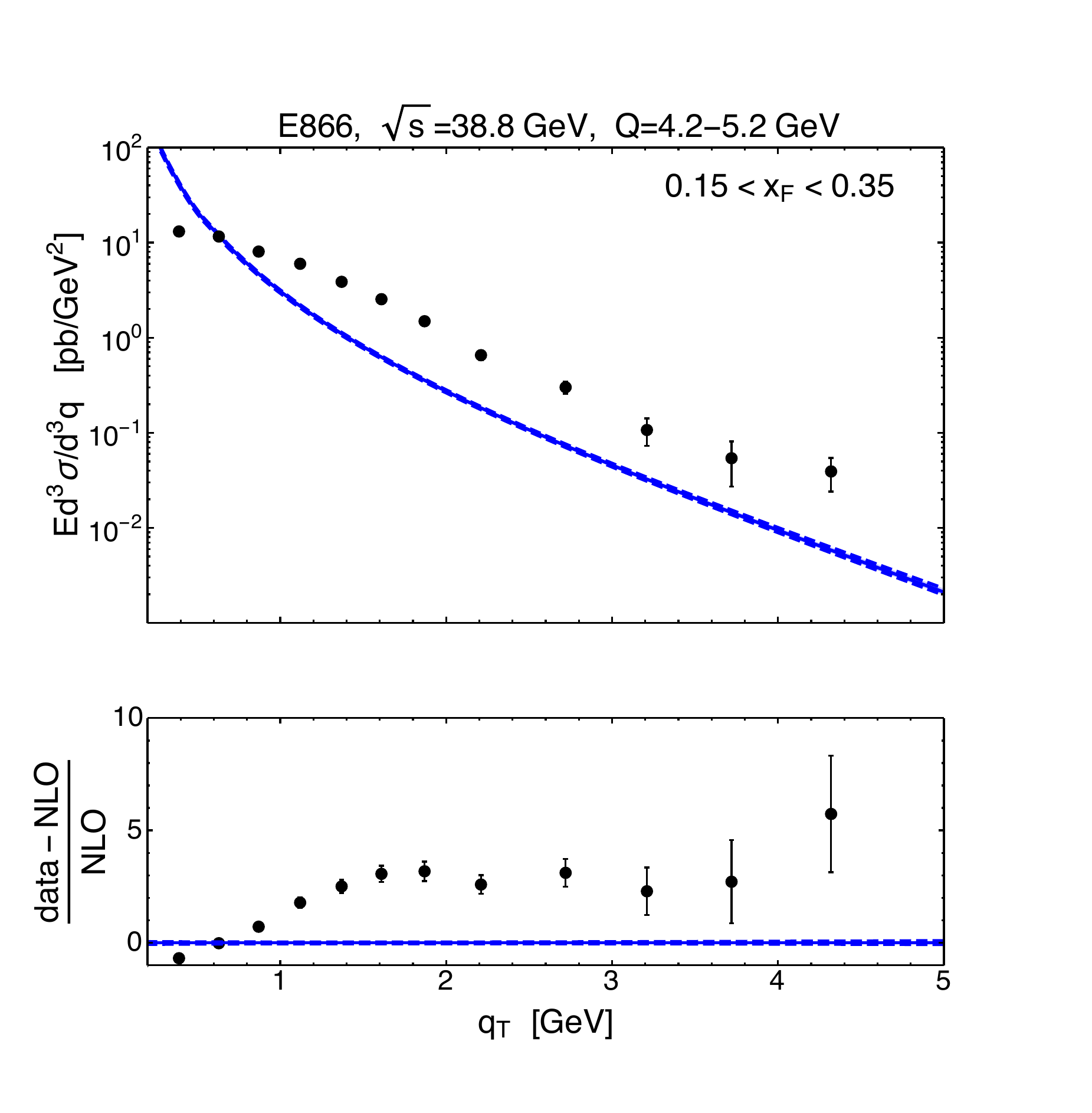} \includegraphics[scale=0.45]{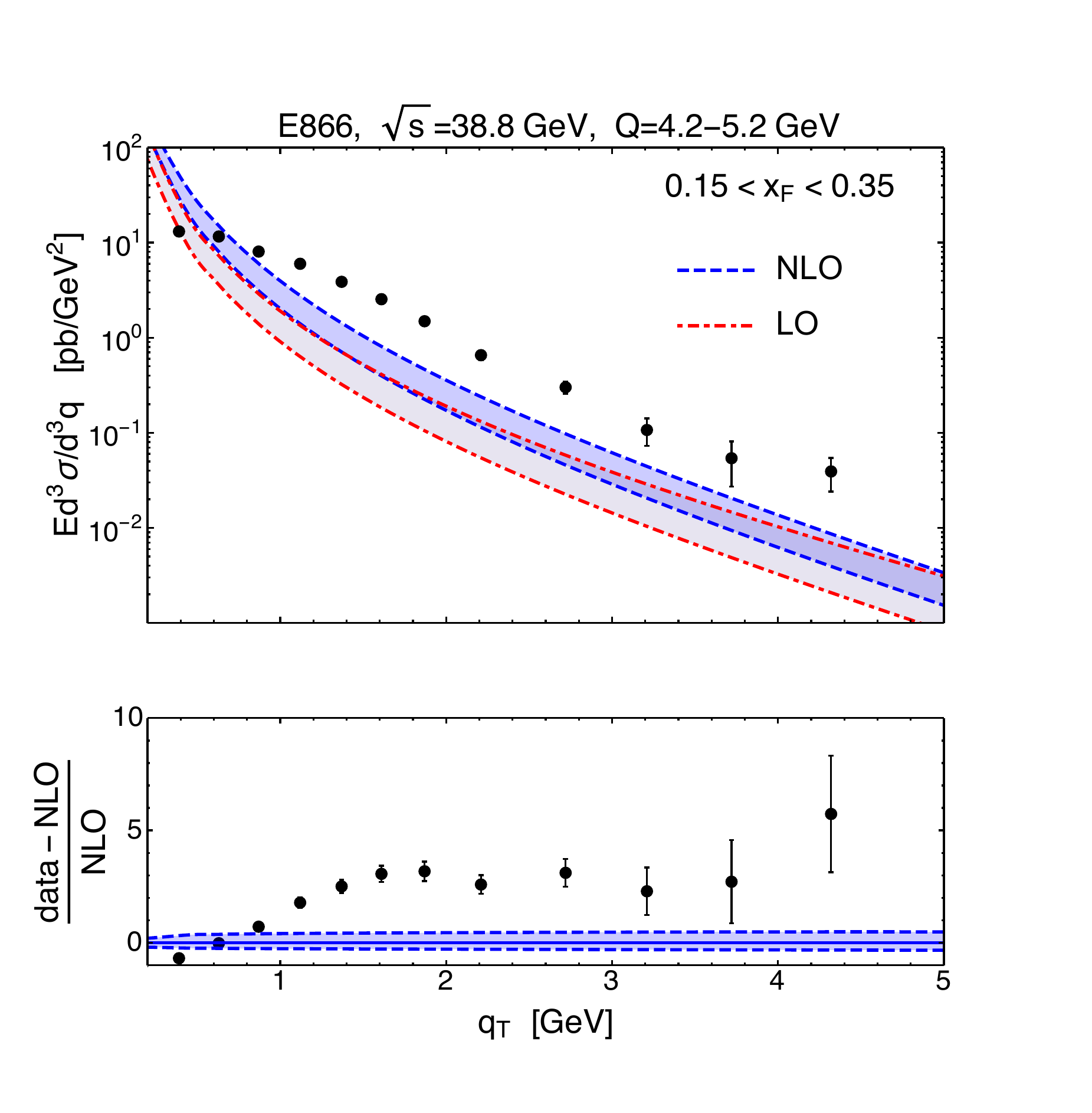}
\end{centering}
\caption{\label{E866_1}Transverse-momentum distribution of Drell--Yan di-muon pairs at $\sqrt s$ = 38.8 GeV in a selected invariant mass range and Feynman-$x$ range: experimental data from Fermilab E866 (hydrogen target) \cite{Webb:2003bj} compared to LO QCD and NLO QCD results. \textbf{Left}: NLO QCD $\left(\mathcal{O}\left(\alpha_{s}^{2}\right)\right)$ calculation with central values of the scales $\mu_{R}=\mu_{F}=Q$ = 4.7 GeV, including a 90\% confidence interval from the CT14 PDF set \cite{Dulat:2015mca}. \textbf{Right}: LO QCD and NLO QCD theoretical uncertainty bands obtained by varying the renormalization and factorization scales independently in the range $Q/2<\mu_{R},\mu_{F}<2Q.$}
\end{figure}

The comparison of our LO and NLO theoretical calculations with the experimental data is shown in Fig.~\ref{E866_1} for the bin 0.15 $\leq x_F \leq$ 0.35 and for the invariant mass range 4.2 GeV $\leq Q \leq$ 5.2 GeV. The lower part of the plot shows the ratio (data-theory)/theory. The error margins of the data points correspond to the sum in quadrature of statistical and systematic uncertainties, including also an overall normalization uncertainty of 6.5\%, as indicated in \cite{Webb:2003bj}. Our theoretical predictions are computed at the average $Q$ value and $ x_F $ of each bin ($Q$ = 4.7 GeV and  $ x_F $ = 0.25 in the case of Fig.~\ref{E866_1}). The left plot of Fig.~\ref{E866_1} shows the comparison of the experimental data with NLO QCD $\left(\mathcal{O}\left(\alpha_{s}^{2}\right)\right)$ predictions for central values of the factorization and renormalization scales, $\mu_{R}=\mu_{F}=Q$. The 90\% confidence interval of the CT14 PDF set \cite{Dulat:2015mca} is included in the plot, but the corresponding variation is barely visible. 

An immediate observation from Fig.~\ref{E866_1} is that the NLO cross section is below the E866 data at high
transverse momenta, $q_T\gtrsim 3$~GeV, even within the relatively large uncertainties that the data have here. 
The NLO cross section falls below the data even much more severely at lower $q_T$ closer to the ``matching regime'' 
with TMD physics, where the experimental uncertainties are much smaller. This provides further evidence to our 
observation above that this regime is presently not well understood theoretically. 
At the same time we emphasize that data from \cite{Webb:2003bj}, {\it integrated over $q_T$}, are in good agreement with theoretical predictions and are commonly used in global PDF fits \cite{Martin:2009iq,Ball:2017nwa} (see, for instance, Section 5.1 of \cite{Webb:2003bj}, where the only relevant discrepancy concerns the lowest mass point ($\left\langle Q\right\rangle \simeq4.4$ GeV) for $0.05<x_{F}<0.25$ (Figs. 5.1-5.5)). 
This suggests that TMD physics may be the main player for the cross section up to relatively high $q_T$,
since the tail at very large $q_T$ makes only a small contribution to the cross section. 

The right plot of Fig.~\ref{E866_1} shows the effect of varying the renormalization and factorization scales independently in the range $Q/2<\mu_{R},\mu_{F}<2Q$, both for the LO QCD $\left(\mathcal{O}\left(\alpha_{s}\right)\right)$ and the NLO QCD $\left(\mathcal{O}\left(\alpha_{s}^{2}\right)\right)$ calculation. The fact that, for $q_{T}\gtrsim 2.5$ GeV, the NLO uncertainty band overlaps with (and is eventually included in) the LO uncertainty band provides some indication that perturbation theory is well-behaved for this process. On the other hand, we also observe that the NLO scale uncertainty band is only marginally more narrow than the LO one.

We have also considered different PDF choices (CTEQ 10 \cite{Lai:2010vv}, NNPDF 2.3 \cite{Ball:2012cx} and MSTW2008 \cite{Martin:2009iq}), obtaining very similar results: the different curves lie within the uncertainty bands shown in the right plot of Fig.~\ref{E866_1}. Such a mild PDF dependence was expected, since the PDFs are well constrained and have small uncertainties in the $x$-range probed in this process. We conclude that PDF uncertainties (unless they are grossly underestimated by the parameterizations) cannot explain the discrepancy between theory and data
at high $q_T$.

\begin{figure}
\begin{centering}
\includegraphics[scale=0.35]{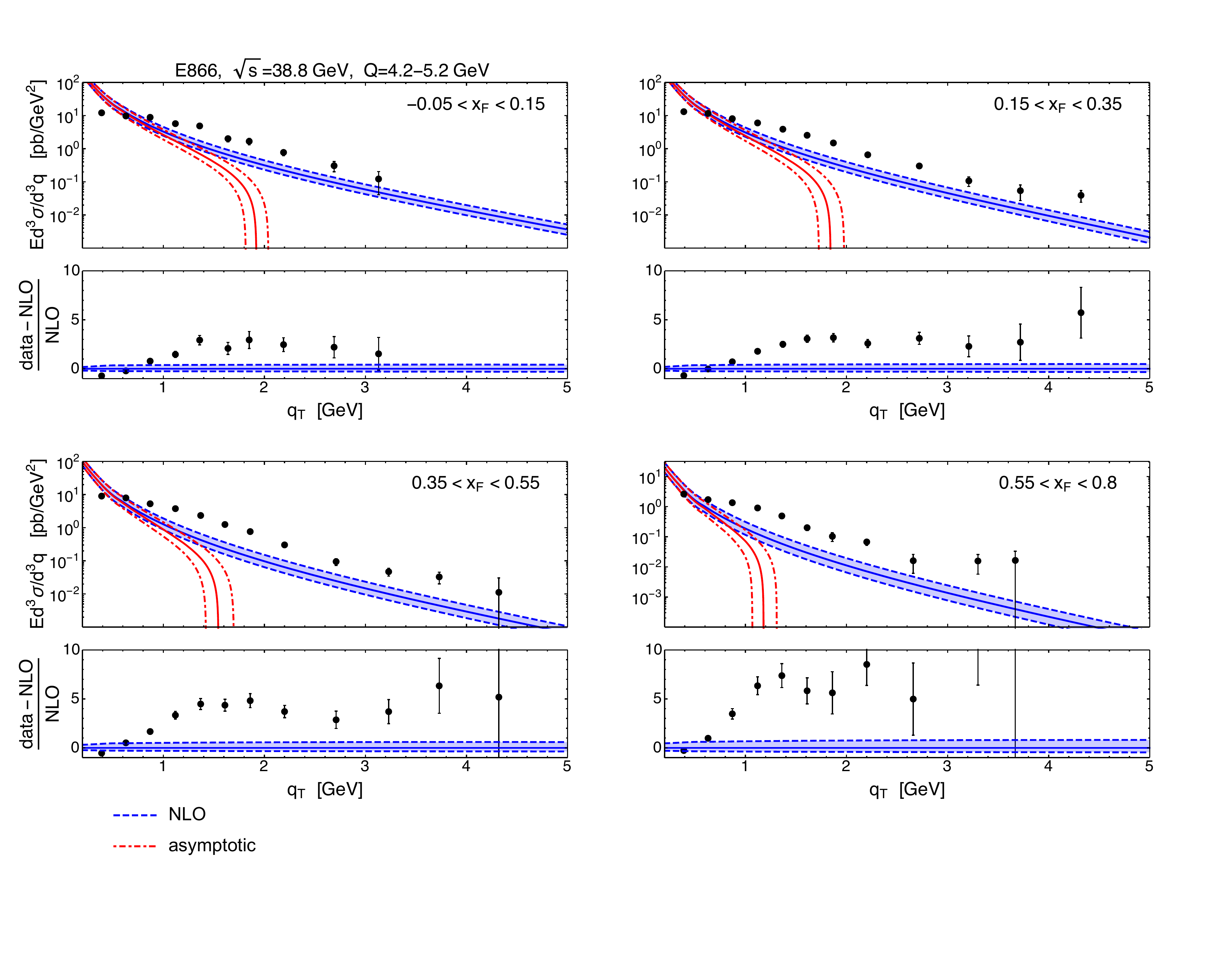}
\end{centering}
\caption{\label{E866_2} E866: comparison between experimental data and NLO QCD predictions for different $x_{F}$ bins. 
We also show the low-$q_T$ asymptotic part of the cross section. For details, see text.}
\end{figure}

\begin{figure}
\begin{centering}
\includegraphics[scale=0.35]{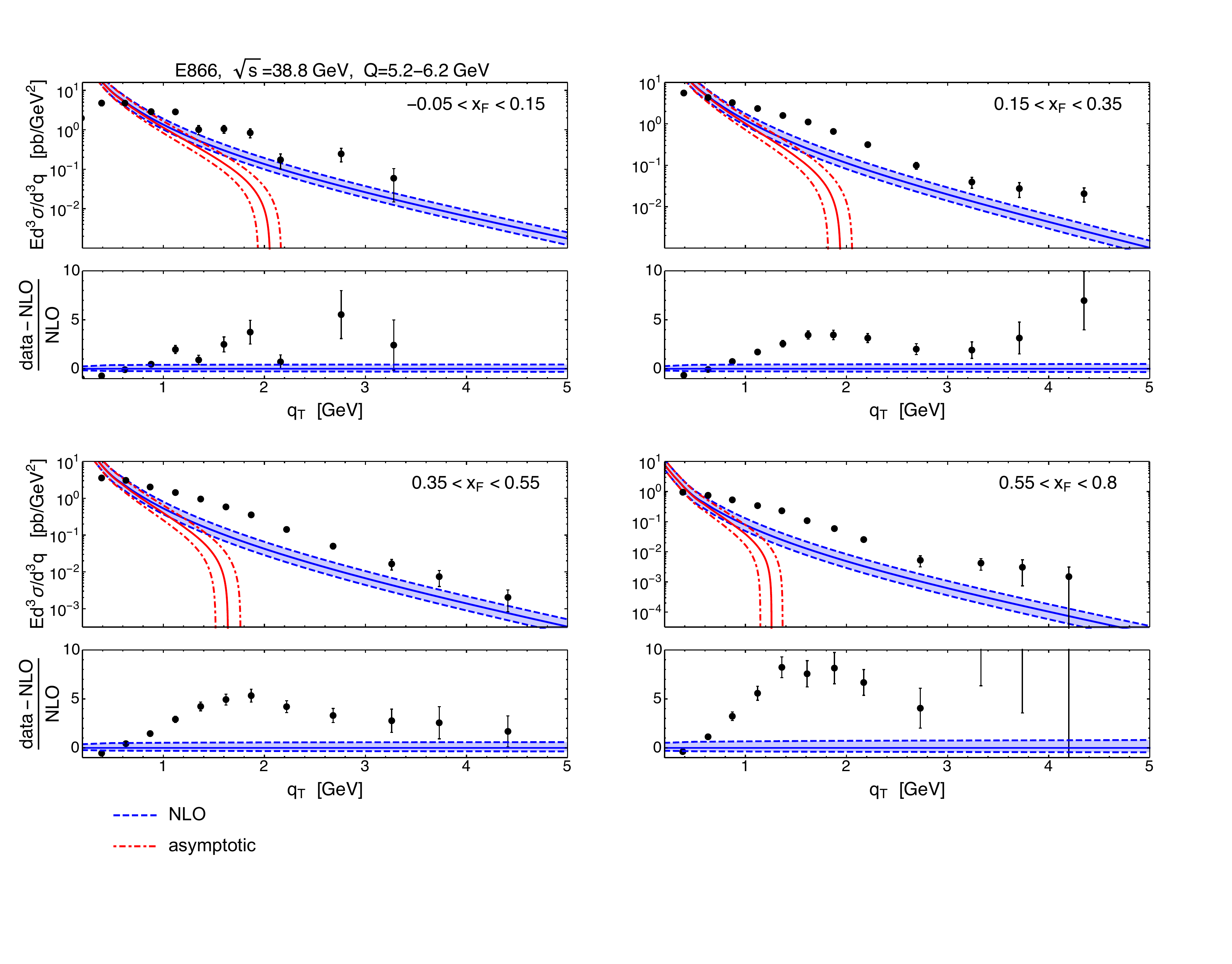}
\end{centering}
\caption{\label{E866_3} E866: comparison between experimental data and NLO QCD predictions for different invariant mass bins. 
We also show the low-$q_T$ asymptotic part of the cross section. For details, see text.}
\end{figure}

The comparison between data and theory for other $x_{F}$ bins (Fig.~\ref{E866_2}) and for a different invariant mass range (Fig.~\ref{E866_3}) gives the same qualitative results. The upper part of each plot contains the NLO QCD $\left(\mathcal{O}\left(\alpha_{s}^{2}\right)\right)$ prediction (blue) 
with its uncertainty band obtained through the customary scale variation $\left(Q/2<\mu_{R},\mu_{F}<2Q\right)$ around the central value $Q$ of the invariant mass range. The lower part of each plot again shows the ratio (data-theory)/theory. 
We also plot the asymptotic expansion of the resummed calculation (red lines). 
The asymptotic result coincides with the fixed order prediction in the region of very low transverse momenta, but it becomes very small (and eventually negative) with increasing $q_{T}$. We show the asymptotic piece in order to obtain a rough guide concerning 
the region where the fixed-order calculation may start to become reliable
\cite{Arnold:1990yk}: ideally,
when $q_T$ is large enough that the difference between the fixed-order and asymptotic calculations (the so-called ``$Y$ term") exceeds the full (``$W+Y$") cross section, one should switch from $W+Y$ to the fixed-order result to obtain more reliable predictions. 
This occurs for $q_T$ values around 1-2 GeV in the present case.
Figures~\ref{E866_2} and~\ref{E866_3} show the same qualitative features seen above: the overall agreement between theory and 
high-$q_T$ data is poor. 
In general, the disagreement between data and theoretical predictions seems to become worse with increasing Feynman-$x_F$ and to be only mildly dependent on the invariant mass $Q$ of the lepton pair. 


\subsection*{R209}

The R209 experiment \cite{Antreasyan:1980yb,Antreasyan:1981uv} (two proton beams colliding at a center-of-mass energy of $\sqrt{s}=62$ GeV) was carried out at the CERN ISR (Intersecting Storage Rings) to search for new particles and test scaling models. The differential cross section $d\sigma/dq_T^{2}$ for the production of a muon pair with transverse-momentum $q_T$ is reported in \cite{Antreasyan:1981eg} for the invariant mass range 5 GeV $< Q <$ 8 GeV. The low transverse momentum part of these data has been included in extractions of TMDs \cite{Konychev:2005iy,Landry:2002ix}. Studies of the whole $q_T$ spectrum can be found in \cite{Gavin:1995ch,Szczurek:2008ga}.  

\begin{figure}
\centering
\includegraphics[scale=0.45]{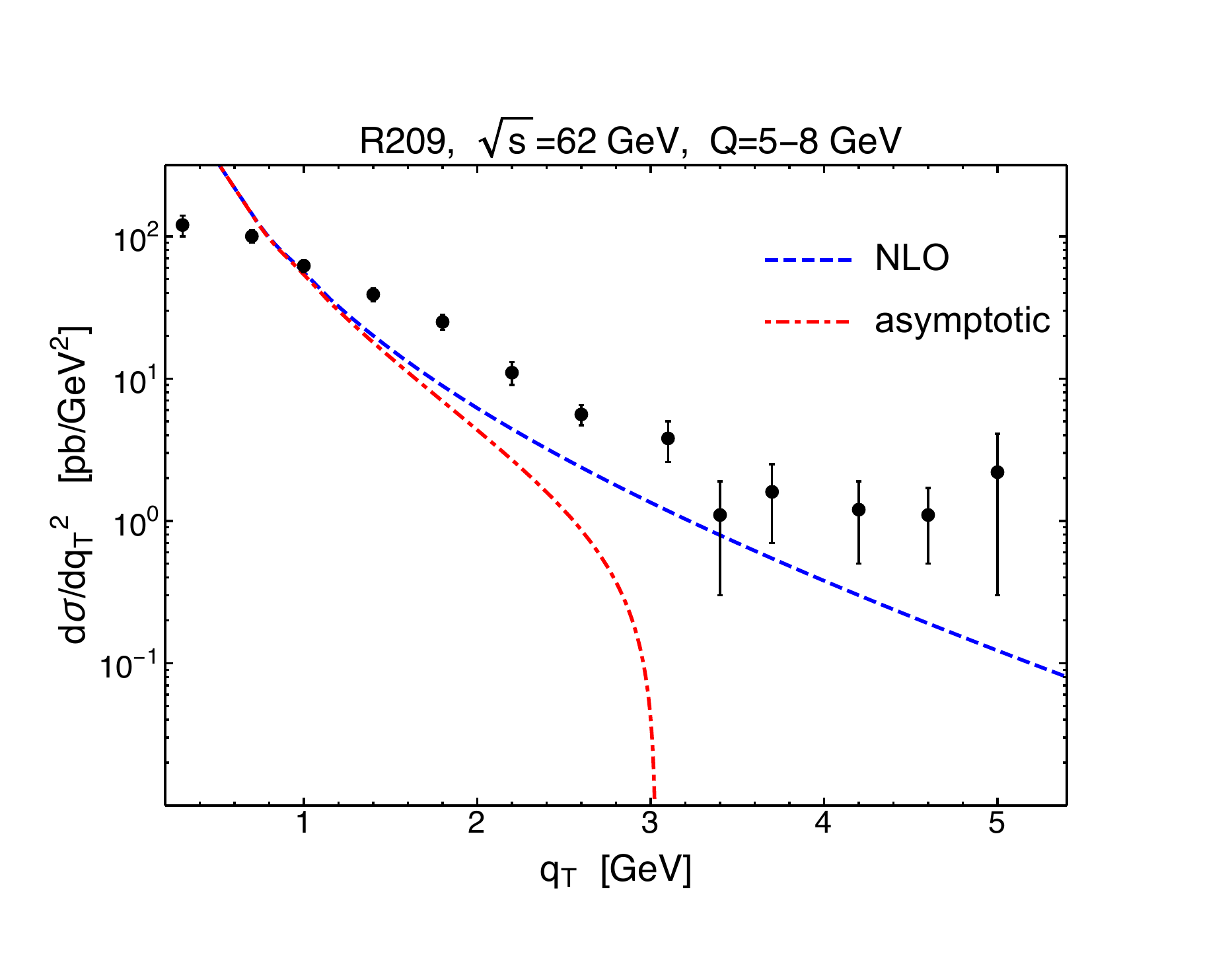}\includegraphics[scale=0.45]{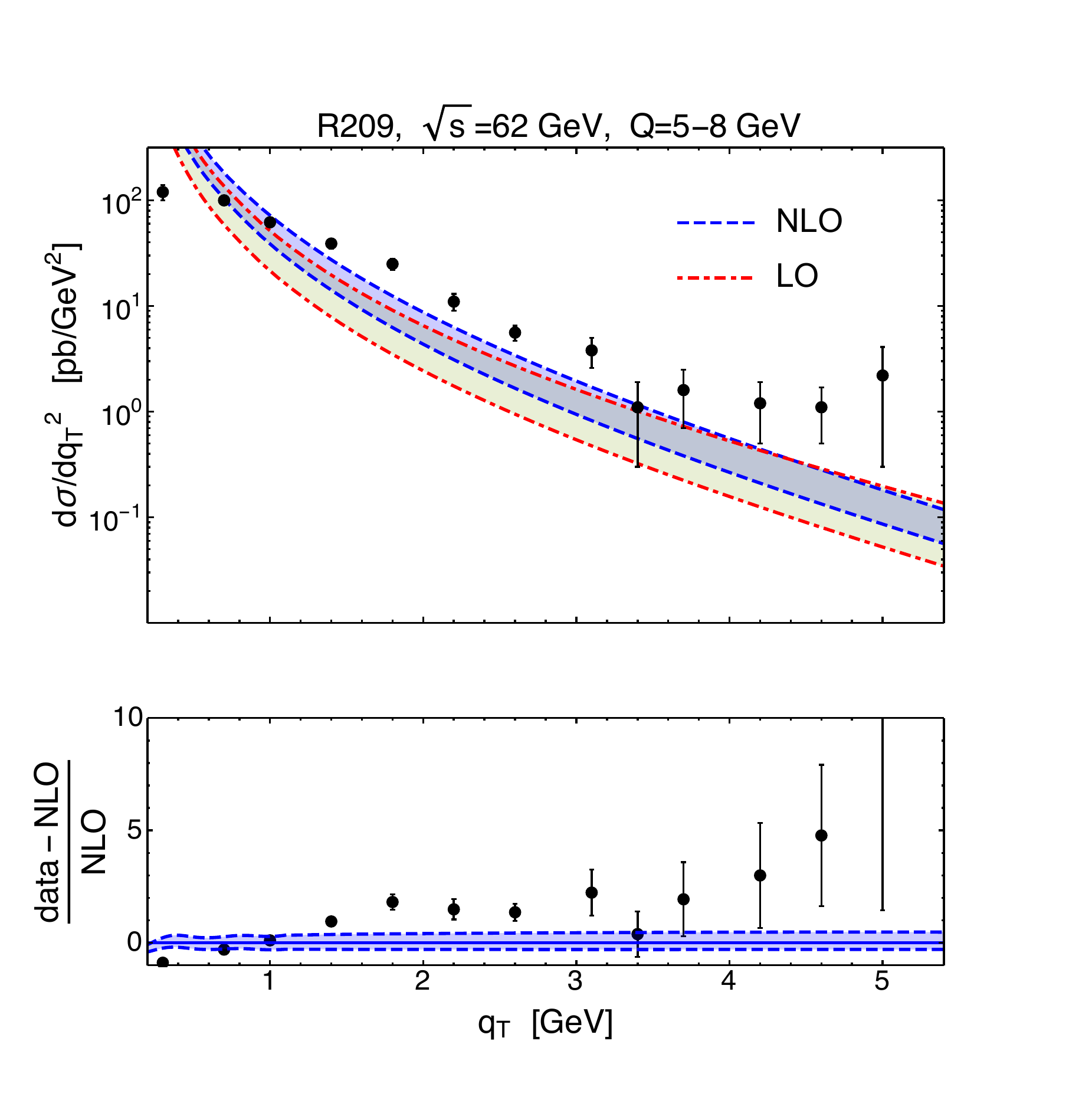}
\caption{\label{fig:R209-data-}\textbf{Left}: R209 data \cite{Antreasyan:1981eg} compared to NLO QCD $\left(\mathcal{O}\left(\alpha_{s}^{2}\right)\right)$. The dashed line shows the asymptotic part. Theoretical results are integrated over the $Q$ range. We have chosen $\mu_{R}=\mu_{F}=Q$. \textbf{Right}: scale variations $\left(Q/2<\mu_{R},\mu_{F}<2Q\right)$ at LO and NLO.}
\end{figure}

Comparisons of our NLO results to the R209 data are shown in Fig.~\ref{fig:R209-data-}. Again NLO is below the data at high $q_T$, 
although the discrepancy is not as statistically significant in this case as for the E866 data. We note that a similar gap between data and theory was reported in \cite{Szczurek:2008ga} in the context of a LO calculation. There, the so-called ``$k_{T}$-factorization'' formalism was claimed to account for the discrepancy. In contrast, in \cite{Gavin:1995ch} the $W+Y$ formalism was reported to match the data over the whole $q_T$ range. 

\subsection*{E288}

\begin{figure}
\begin{centering}
\includegraphics[scale=0.35]{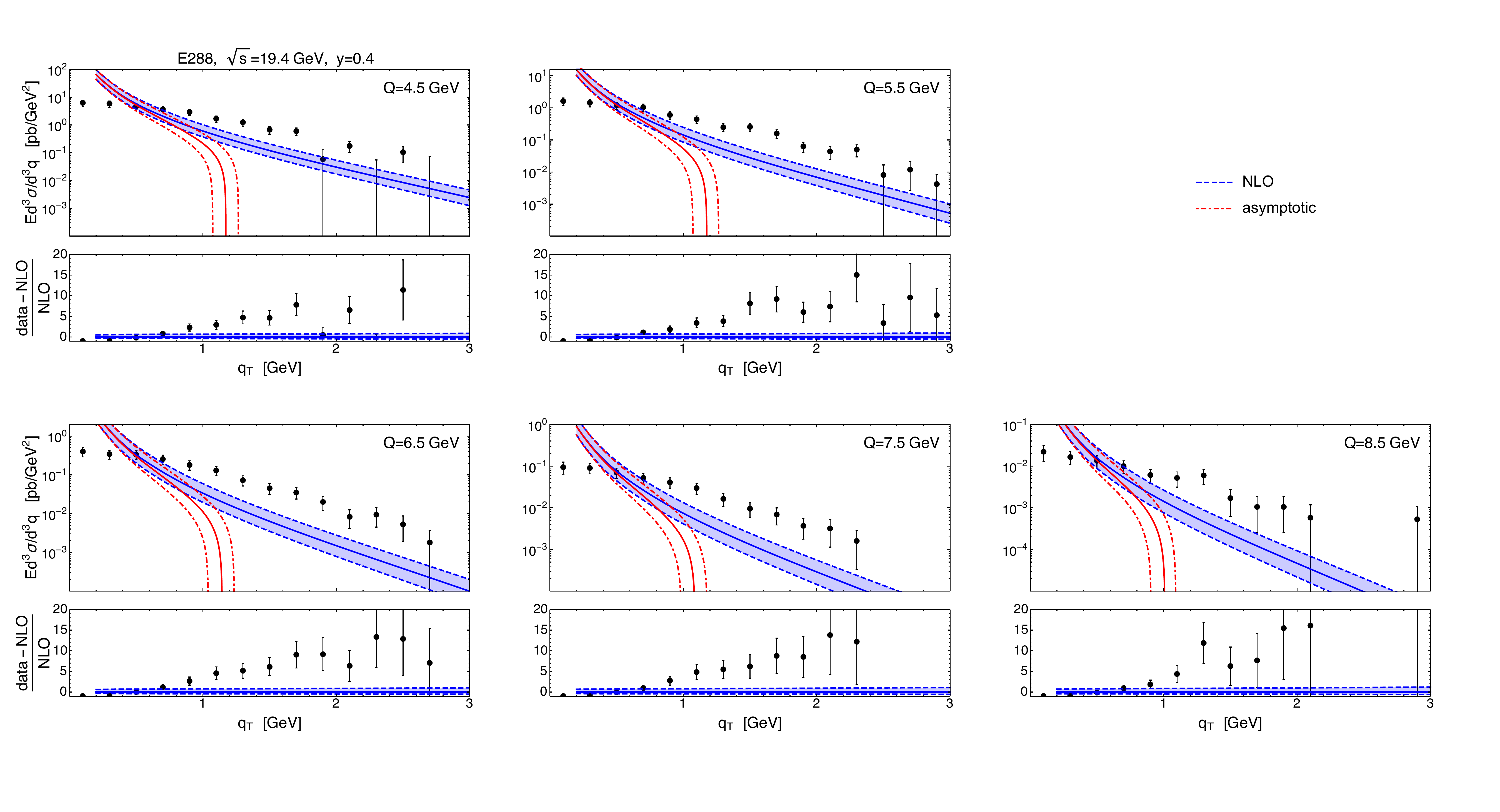}
\end{centering}
\caption{\label{E288_200} E288: experimental data vs. NLO QCD predictions for $\eta$=0.4 and different invariant mass bins.}
\end{figure}

\begin{figure}
\begin{centering}
\includegraphics[scale=0.35]{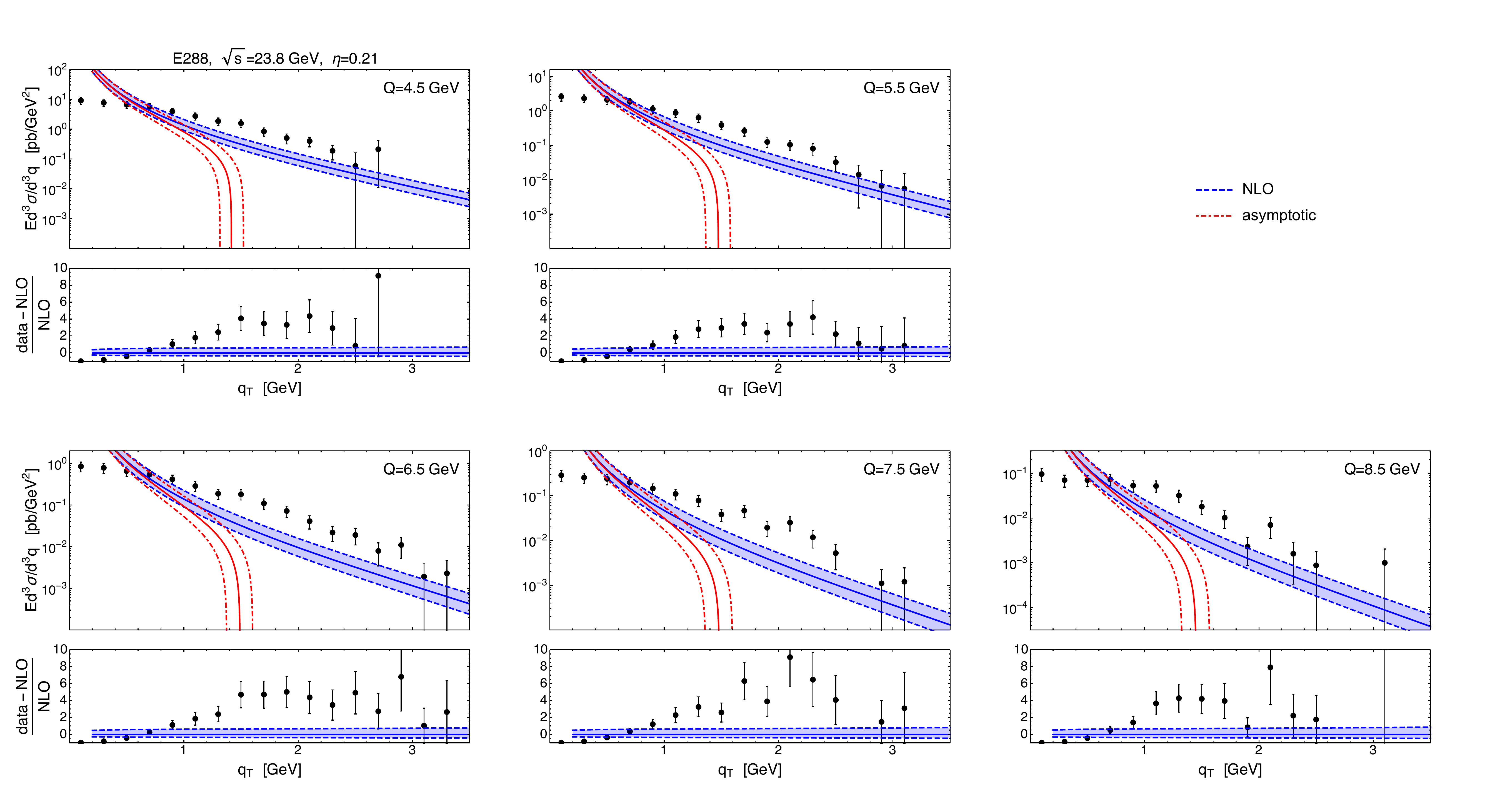}
\end{centering}
\caption{\label{E288_300} E288: experimental data vs. NLO QCD predictions for $\eta$=0.21 and different invariant mass bins.}
\end{figure}

\begin{figure}
\begin{centering}
\includegraphics[scale=0.35]{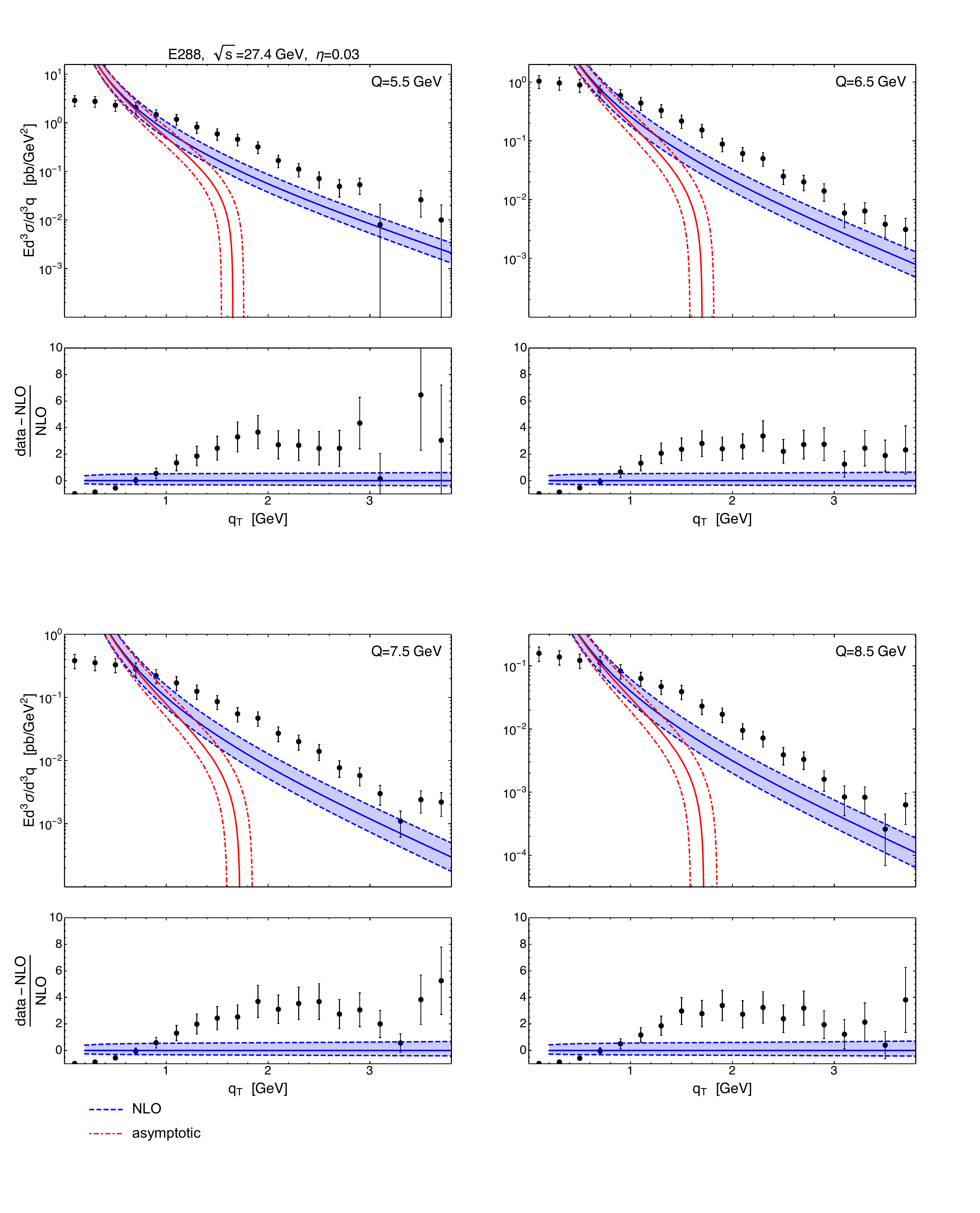}
\end{centering}
\caption{\label{E288_400A} Additional plots for E288: experimental data vs. NLO QCD predictions for $\eta$=0.03 and different invariant mass bins.}
\end{figure}

\begin{figure}
\begin{centering}
\includegraphics[scale=0.35]{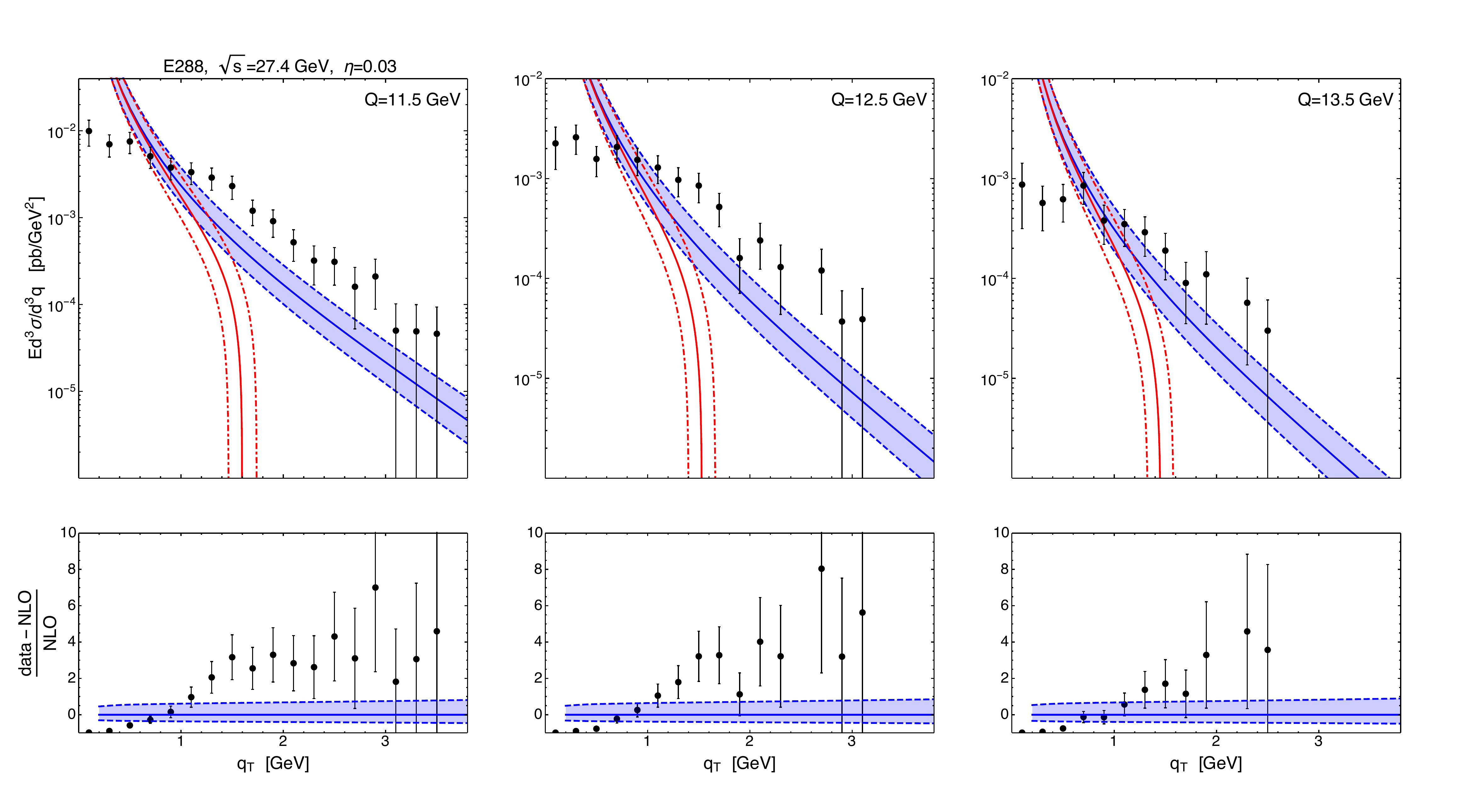}
\end{centering}
\caption{\label{E288_400B} Additional plots for E288: experimental data vs. NLO QCD predictions for $\eta$=0.03 and different invariant mass bins.}
\end{figure}

The E288 experiment~\cite{Ito:1980ev} measured the invariant cross section $E d^{3}\sigma / d^3q$, at fixed photon rapidity, for the production of $\mu^+ \mu^-$ pairs in the collision of a proton beam with a fixed target composed of either Cu or Pt. The measurements were performed using proton incident energies of $200$, $300$ and $400$ GeV, producing three different data sets. The respective center of mass energies are $\sqrt{s}=19.4,23.8,27.4$ GeV.  Our results are shown in Figs.~\ref{E288_200},\;\ref{E288_300},\;\ref{E288_400A},\;\ref{E288_400B}. The comparison to data shows the same features as before. We have tested the importance of nuclear effects by computing the cross sections also with the nCTEQ15 \cite{Kovarik:2015cma} and CT14 \cite{Dulat:2015mca} nuclear PDFs. These turn out to lead to almost indistinguishable results.  We note that the low transverse momentum part of the E288 data has been used for extractions of TMDs \cite{Konychev:2005iy,Landry:2002ix,DAlesio:2014mrz,Bacchetta:2017gcc,Scimemi:2017etj}. 


\subsection*{E605}

\begin{figure}
\begin{centering}
\includegraphics[scale=0.35]{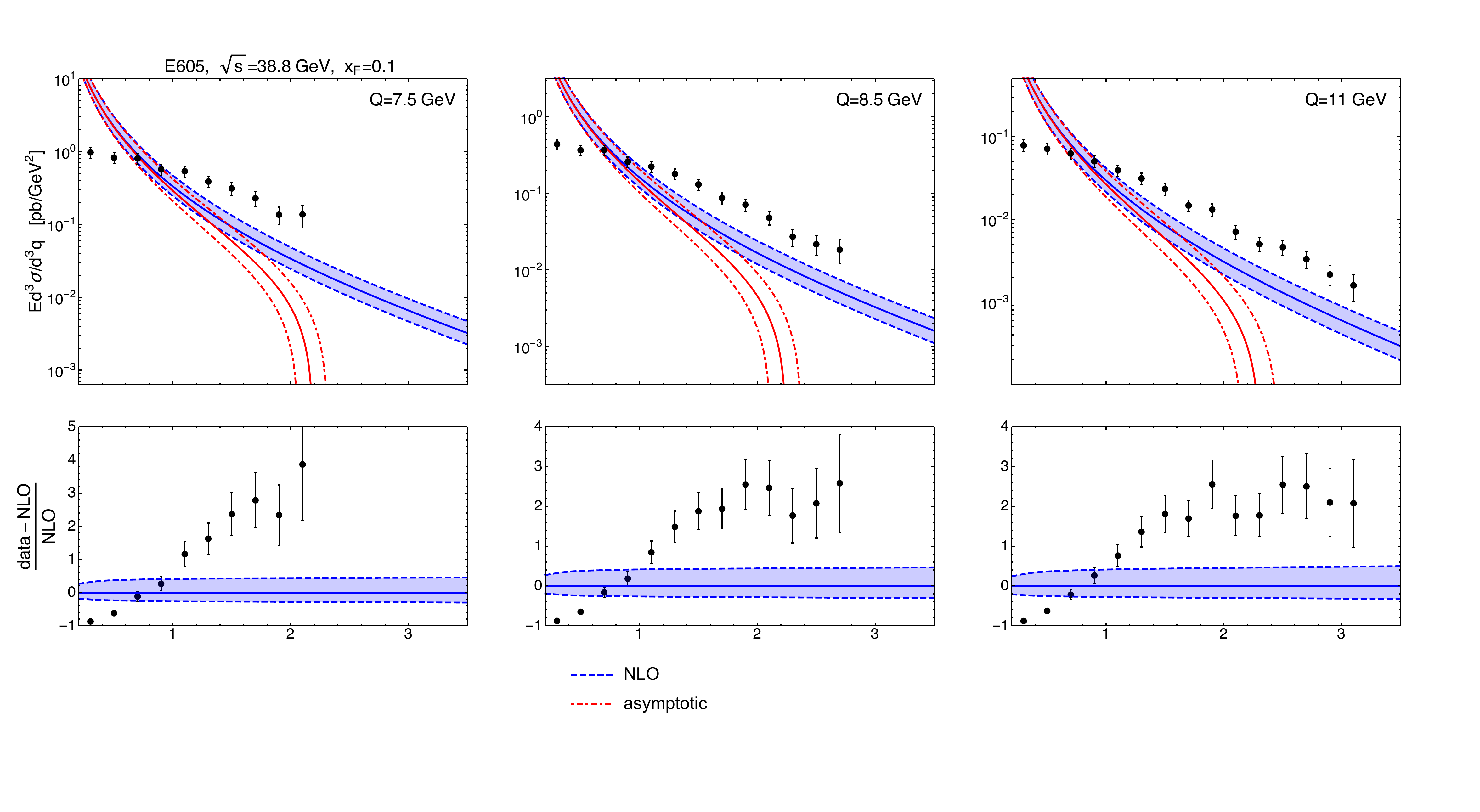}
\end{centering}
\caption{\label{E605} E605: experimental data vs. NLO QCD predictions for $x_{F}$=0.1 and different invariant mass bins.}
\end{figure}

We also consider the set of measurements of $E d^{3}\sigma/d^3q$ in the E605~\cite{Moreno:1990sf} experiment, extracted from an 800-GeV proton beam incident on a copper fixed target ($\sqrt{s}=38.8$ GeV). Results at fixed $x_F=0.1$ are shown in Fig.~\ref{E605}. The low transverse momentum part of these data has also been included in extractions of TMDs \cite{Konychev:2005iy,Landry:2002ix,Bacchetta:2017gcc}. 

\subsection*{PHENIX}

\begin{figure}
\begin{centering}
\includegraphics[scale=0.35]{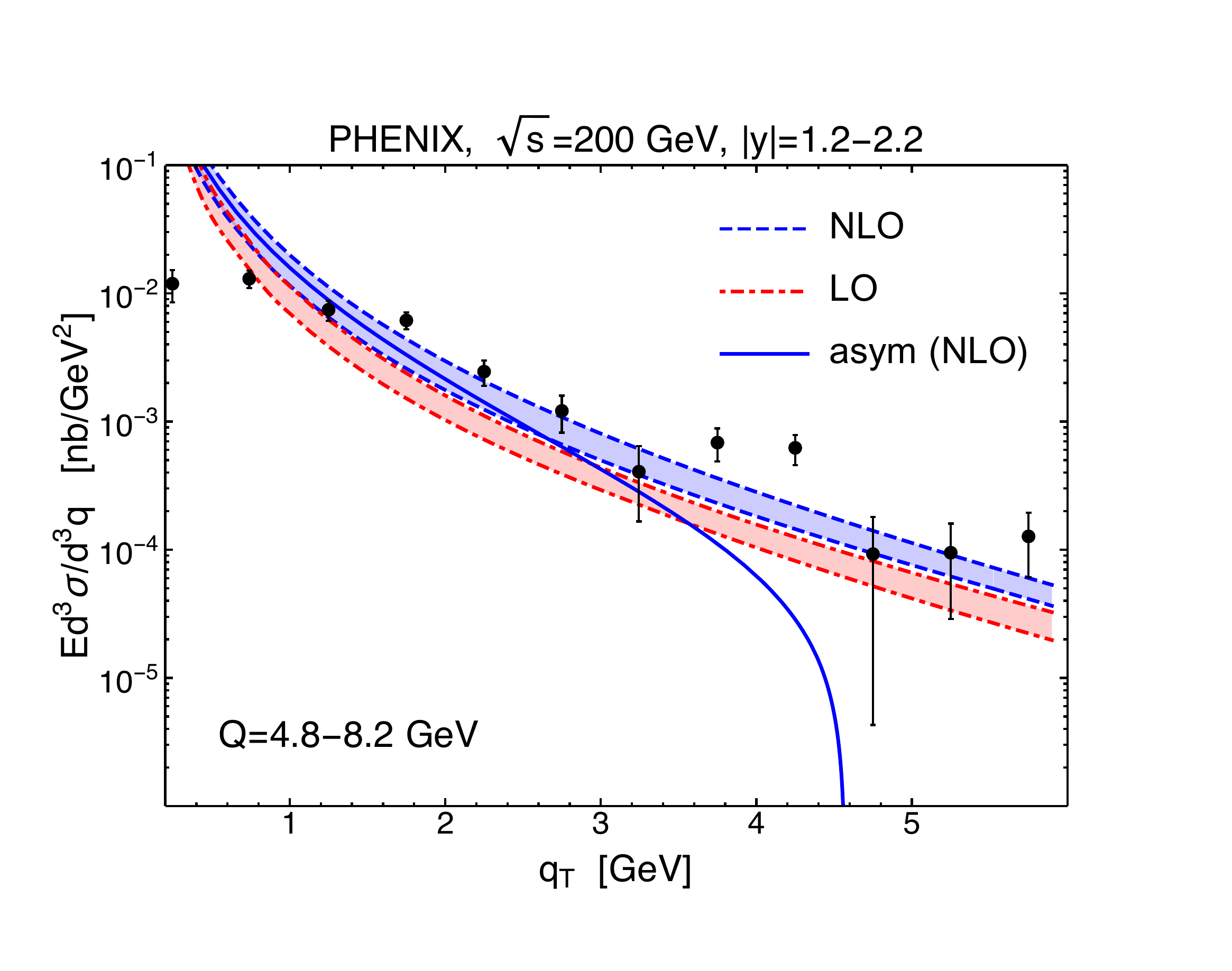}
\end{centering}
\caption{\label{PHENIX} PHENIX: experimental data vs. NLO QCD predictions for 1.1$<\left|y\right|<$2.2 and 4.8 GeV $<Q<$ 8.2 GeV.}
\end{figure}

Finally, we also compare to the recent measurement~\cite{Aidala:2018ajl} performed by the PHENIX collaboration at the Relativistic Heavy Ion Collider in $pp$ collisions at $\sqrt{s}$=200 GeV. The experimental points are taken from Fig.~33 of~\cite{Aidala:2018ajl} and compared to LO QCD and NLO QCD, including theoretical uncertainties, in Fig.~\ref{PHENIX}. The asymptotic expansion of the $W$ term to NLO is also shown. Evidently, the comparison between NLO and the data is overall satisfactory in this case. It thus appears that there is a qualitative difference between the fixed-target and collider regimes.

\section{Threshold resummation}

As we have seen in~Fig.~\ref{E866_1}, the NLO corrections to the $q_T$-differential cross sections are quite sizable. It is therefore important to investigate in how far beyond-NLO perturbative corrections might be relevant for obtaining a better agreement with the data. For the kinematics relevant for the Fermilab and CERN experiments, the invariant mass and transverse momentum of the Drell--Yan pair are such that the production is relatively close to partonic threshold, where a new class of logarithms (separate from that mentioned above at low $q_T$) arises. The summation of these logarithms to all orders is known as {\it threshold resummation}. We note that large corrections from threshold resummation have been found previously in purely hadronic single-inclusive processes such as $pp\to \pi X$~\cite{deFlorian:2005yj,Hinderer:2018nkb}, which motivates a corresponding study for the high-$q_{T}$ Drell--Yan cross section $pp\to \gamma^*X\to \ell^+\ell^-X$ that will be carried out in this section. The relevant formalism has been developed in Refs.~\cite{Kidonakis:1999ur,Kidonakis:2003xm,Gonsalves:2005ng,Kidonakis:2014zva,Muselli:2017bad}, although in most of these papers only fixed-order (NNLO) expansions of the resummed cross sections have been considered, and in~\cite{deFlorian:2005fzc} for the closely related high-$q_T$ Higgs production cross section. We follow here the approach taken in the latter reference. 

\subsection*{Factorized cross section and Mellin moments} 

For simplicity we will focus here just on the transverse momentum distribution of the lepton pair and integrate over the full range of allowed rapidities of the virtual photon. We therefore consider
\be
\f{ d\sigma}{dQ^2dq_T^2 } \,=\, \int_{y^-}^{y^+} dy
 \f{ d\sigma}{dQ^2dq_T^2 dy} \, ,
\ee
where 
\be\label{kinlim}
y^+=-y^-=\f{1}{2} \ln\f{1+\sqrt{1-4 s\, \mT^2/(s+Q^2)^2}}{1-\sqrt{1-4 s\, \mT^2/(s+Q^2)^2}}\;,
\ee
with $\mT\equiv\sqrt{Q^2+\pt^2}$ the transverse mass. The integrated differential cross section for 
$h_1h_2\to \ell^-\ell^+X$ may be written in factorized form as
\beeq
\label{eq:1}
\f{ d\sigma}{dQ^2dq_T^2} &=& \sum_{a,b}\, 
\int_0^1 dx_1 \, f_{a/h_1}\left(x_1,\mu_{F}^2\right) \,
\int_0^1 dx_2 \, f_{b/h_2}\left(x_2,\mu_{F}^2\right) \, 
\f{d\hat{\sigma}_{ab}}{{dQ^2dq_T^2}}\nn\\[2mm]
&\equiv&\frac{\sigma_0}{q_T^2 Q^2} \sum_{a,b}
\int_{y_T^2}^1 dx_1 \, f_{a/h_1}\left(x_1,\mu_{F}^2\right)
\int_{y_T^2/x_1}^1 dx_2 \, f_{b/h_2}\left(x_2,\mu_{F}^2\right) 
\omega_{ab}\left(\yTh,r,\frac{\mu_F^2}{Q^2},
\frac{\mu_R^2}{Q^2},\alpha_s(\mu_R^2)\right) \, ,
\eeeq
where $\sig_0=4\pi\alpha^2/(9Q^2)$, $f_{a/h_1}$ and $f_{b/h_2}$ are the PDFs, and where
$\sh = s x_1 x_2$ is the partonic center-of-mass energy squared. In the second line we have written out the variables that the dimensionless hard-scattering functions $\omega_{ab}$ may depend on. It is convenient to write the kinematical arguments of $\omega_{ab}$ as 
\beeq
\yTh&\equiv& \frac{\pt+\mT}{\sqrt{\hat{s}}} \,,\nn\\[2mm]
r &\equiv&\frac{q_T}{\mT}\,.
\eeeq
Note that in terms of these we have $q_T/\sqrt{\hat{s}}=\yTh r/(1+r)$ and $Q^2/\hat{s}=\yTh^2(1-r)/(1+r)$. In~(\ref{eq:1}) we have also introduced the corresponding hadronic variable 
\be
\yT\,\equiv\, \frac{\pt+\mT}{\sqrt s} \,=\,\sqrt{x_1x_2}\,\yTh\;.
\ee 
From Eq.~(\ref{kinlim}) we see that $\yT\leq 1$. Likewise, we also have $\yTh\leq 1$, which immediately leads to the limits on $x_1$ and $x_2$ given in Eq.~(\ref{eq:1}). The rapidity-integrated cross section thus takes the form of a convolution of the hard-scattering functions $\omega_{ab}$ with the PDFs. The perturbative expansion of the $\omega_{ab}$ reads
\be
\omega_{ab}\,=\,\frac{\alpha_s(\mu_R^2)}{\pi}\,\omega_{ab}^{(0)}+{\cal O}(\alpha_s^2)\,.
\ee
There are two LO partonic channels, $q\bar{q}\to \gamma^*g$ and $qg\to \gamma^*q$. As before, $\mu_F$ and $\mu_R$ in~(\ref{eq:1}) denote the factorization and renormalization scales. 

For $\yTh\rar 1$ the partonic center-of-mass energy is just sufficient to produce the lepton pair with mass $Q$ and transverse momentum $q_T$. Therefore, $\yTh=1$ sets a threshold for the process. As is well known~\cite{Sterman:1986aj,Catani:1990rp}, the partonic cross sections receive large logarithmic corrections near this threshold. At the $k$th order of perturbation theory for the $\omega_{ab}$, there are logarithmically enhanced contributions of the form $\as^k \, \ln^m (1-\yTh^2)$, with $m\leq 2k$. These logarithmic terms are  due to soft and/or collinear gluon radiation and dominate the perturbative expansion when the process is kinematically close to the partonic threshold. We note that $\hat{y}_T$ becomes 
especially large when the partonic momentum fractions approach their lower integration limits. Since the PDFs rise steeply towards small argument, this enhances the relevance of the threshold regime, and the soft-gluon effects are relevant even when the hadronic center-of-mass energy is much larger than the produced transverse mass and transverse momentum of the final state. In the following, we discuss the resummation of the large logarithmic corrections to all orders in $\as$.

The resummation of the soft-gluon contributions is carried out in Mellin-$N$ moment space, where the convolutions in Eq.~(\ref{eq:1}) between parton distributions and subprocess cross sections factorize into ordinary products. We take Mellin moments of the hadronic cross section in $y_T^2$:
\be
\int_0^1 d \yT^2 \,{(\yT^2)}^{N-1} \frac{d \sig}{dQ^2d \ptsq} \,=\, \frac{\sigma_0}{q_T^2 Q^2}\sum_{a,b}
\tilde{f}_a (N+1, \mu^2_F) \,\tilde{f}_b (N+1, \mu^2_F)\,\tilde{\omega}_{ab}\left(N,r,\frac{\mu_F^2}{Q^2},
\frac{\mu_R^2}{Q^2},\alpha_s(\mu_R^2)\right)\,,
\ee
where the corresponding moments of the partonic hard-scattering functions are
\be
\tilde{\omega}_{ab}\left(N,r,\frac{\mu_F^2}{Q^2},
\frac{\mu_R^2}{Q^2},\alpha_s(\mu_R^2)\right)\,\equiv\,
\int_0^1 d \yTh^2 \, (\yTh^2)^{N-1}  \,\omega_{ab}\left(\yTh,r,\frac{\mu_F^2}{Q^2},
\frac{\mu_R^2}{Q^2},\alpha_s(\mu_R^2)\right)\,,
\ee
and where the $\tilde{f}_{a,b} (N+1, \mu^2_F)$ are the moments of the parton distributions. The threshold limit $\hat{y}_T^2\to 1$ corresponds to $N\to \infty$, and the leading soft-gluon corrections now arise as terms $\propto \as^k \ln^mN$, $m\leq 2k$. The NLL resummation procedure 
we present here deals with the ``towers'' $\as^k \ln^{m}N$ for $m=2k,2k-1,2k-2$.

\subsection*{Resummation to NLL}
\label{subsec:nll}

In Mellin-moment space, threshold resummation results in exponentiation of the soft-gluon corrections. In case of lepton pair production at high $q_T$, the resummed $y$-integrated cross section reads~\cite{deFlorian:2005fzc}:
\begin{align}
\label{eq:res}
\tilde{\omega}^{{\rm (res)}}_{ab}(N) \,=\,  C_{ab\to \gamma^* c}\,
\Delta^a_{N+1}\, \Delta^b_{N+1}\, J_{N+1}^c
\Delta^{{\rm (int)} ab\rightarrow \gamma^* c}_{N+1} \,
\tilde{\omega}^{(0)}_{ab}(N) \;  ,
\end{align}
where for simplicity we have suppressed the arguments of all functions other than the Mellin variable $N$.  Each of the ``radiative factors'' $\Delta_N^{a,b}$, $J_N^c$, $\Delta^{{\rm (int)} ab\rightarrow \gamma^* c}_{N}$ is an exponential. The factors $\Delta^{a,b}_N$ represent the effects of soft-gluon radiation collinear to initial partons $a$ and $b$. The function $J^{c}_N$ embodies collinear, soft or hard, emission by the 
parton $c$ that recoils against the lepton pair. Large-angle soft-gluon emission is accounted for by the factors $\Delta^{{\rm (int)} ab\rightarrow \gamma^* c}_{N}$, which depend on the partonic process under consideration. Finally, the coefficients $C_{ab\to \gamma^* c}$ contain $N$-independent hard contributions arising from one-loop virtual corrections and non-logarithmic soft corrections. The structure of the resummed expression is similar to that for the large-$q_T$ $W$-boson production cross section \cite{Kidonakis:1999ur,Kidonakis:2003xm,Gonsalves:2005ng,Kidonakis:2014zva} or, in the massless limit, to that for prompt-photon production in hadronic collisions \cite{Catani:1998tm,Catani:1999hs}.

The expressions for the radiative factors are
\ba
\ln \Delta^{a}_N &=&  \int_0^1 dz \,\f{z^{N-1}-1}{1-z}
\int_{\mu_{F}^2}^{(1-z)^2 Q_0^2} \f{dq^2}{q^2} A_a\big(\as(q^2)\big)\, , \nn \\[2mm]
\ln J^{c}_N &=&  \int_0^1 dz \,\f{z^{N-1}-1}{1-z} \Bigg[ \int_{(1-z)^2
    Q_0^2}^{(1-z) Q_0^2} \f{dq^2}{q^2} A_c\big(\as(q^2)\big) +\f{1}{2}
  B_c\big(\as((1-z)Q_0^2)\big) \Bigg]\, , \nn \\[2mm]
\ln\Delta^{{\rm (int)} ab\rightarrow \gamma^* c}_N&=& 
 \int_0^1 dz \,\f{z^{N-1}-1}{1-z}\, D_{ ab\to \gamma^* c}\big(\as((1-z)^2 Q_0^2)\big)\, ,
\ea
where $Q_0^2\equiv q_T(q_T+m_T)$. Each of the coefficients ${\mathcal C}= A_a,\, B_a,\,D_{ ab\to \gamma^* c}$ is a power series in the coupling constant $\as$, ${\mathcal C}= \sum_{i=1}^{\infty}(\as / \pi)^i{\mathcal C}^{(i)}$. The universal LL and NLL coefficients $A_a^{(1)}$, $A_a^{(2)}$ and  $B_a^{(1)}$ are well known \cite{Kodaira:1982az,Catani:1988vd}:
\begin{equation} 
\label{A12coef} 
A_a^{(1)}= C_a
\;,\;\;\;\; A_a^{(2)}=\frac{1}{2} \, C_a K \;,\;\;\;\; B_a^{(1)}=\gamma_a\,,
\end{equation} 
with
\begin{equation} 
\label{kcoef} 
K = C_A \left( \frac{67}{18} - \frac{\pi^2}{6} \right)  
- \frac{5}{9} N_f \;, 
\end{equation}
where $C_g=C_A=N_c=3$, $C_q=C_F=(N_c^2-1)/2N_c=4/3$, $\gamma_q=-3/2 C_F$ and $\gamma_g=-(11 C_A - 2 N_f )/6$. The process-dependent coefficients $D_{ ab\to \gamma^* c}$ may be obtained as for the Higgs production cross section considered in~\cite{deFlorian:2005fzc}: 
\ba
D_{ ab\to \gamma^* c}\,=\, (C_a +C_b -C_c) \ln\frac{1+r}{r}\, .
\ea
The coefficient is evidently just proportional to a combination of the color factors for each hard parton participating in the process. This simplicity is due to the fact that there is just one color structure for a process with only three external partons. 

The final ingredients for the resummed cross section in~(\ref{eq:res}) are the lowest-order partonic cross sections in Mellin-moment space, $\tilde{\omega}^{(0)}_{ab}(N)$, and the coefficients $C_{ab\to \gamma^* c}$. The expressions for the former are presented in Appendix~\ref{app:lo}.  
At NLL accuracy, we only need to know the first-order terms in the expansion $C_{ab\to \gamma^* c}= 1+\sum_{i=1}^{\infty} (\as / \pi)^i C_{ab\to \gamma^* c}^{(i)}$. These coefficients may be obtained by comparison to the full NLO results given in Ref.~\cite{Gonsalves:1989ar} (see also~\cite{Kidonakis:2014zva}), after transforming to moment space. Our explicit results for the one-loop coefficients $C_{ab\to \gamma^* c}^{(1)}$ are given in Appendix~\ref{app:coeff}.   

In order to organize the resummation according to the logarithmic accuracy of the Sudakov exponents we expand the latter to NLL as~\cite{Catani:1996yz}
\ba
\label{lndeltams}
\!\!\! 
\ln \Delta_N^a(\as(\mu_R^2),Q_0^2/\mu_R^2;Q_0^2/\mu_{F}^2) 
&\!\!=\!\!& \ln \bar{N} \;h_a^{(1)}(\lambda) +
h_a^{(2)}(\lambda,Q_0^2/\mu_R^2;Q_0^2/\mu_{F}^2)  \,, \nn \\[2mm]
\label{lnjfun}
\ln J_N^a(\as(\mu_R^2),Q_0^2/\mu_R^2) &\!\!=\!\!& \ln \bar{N} \;f_a^{(1)}(\lambda) +
f_a^{(2)}(\lambda,Q_0^2/\mu_R^2) \,, \nn \\[2mm]
\label{lnintfun}
\ln\Delta^{(int) ab\rightarrow \gamma^* c}_{N}(\as(\mu_R^2))
&\!\!=\!\!& \frac{D_{ab\rightarrow \gamma^* c}^{(1)}}{2\pi b_0} \;\ln (1-2\lambda)  \,,
\ea
with $\lambda=\b0 \as(\mu^2_R) \ln \bar{N}$. Here, $\bar{N}=N{\mathrm{e}}^{\gamma_E}$ where $\gamma_E$ is the Euler constant. The LL and NLL functions $h^{(1,2)}$ and $f^{(1,2)}$ are given in Appendix~\ref{app:hf}.

\subsection*{Matching and inverse Mellin transform}
\label{subsec:match}

When performing the resummation, one wants to make full use of the available fixed-order cross section, which in our case is NLO (${\cal O}(\alpha_s^2)$). Therefore, one matches the resummed result to the fixed-order expression. This is achieved by expanding the resummed cross section to ${\cal O}(\as^2)$, subtracting the expanded result from the resummed one, and adding the full NLO cross section:
\beeq
\label{hires}
\f{d\sigma^{\rm (match)}}{dQ^2dq_T^2} &=& \sum_{a,b}\,
\;\int_{C_{MP}-i\infty}^{C_{MP}+i\infty}
\;\frac{dN}{2\pi i} \;\left( y_T^2 \right)^{-N}
\; f_{a/h_1}(N+1,\mu_F^2) \; f_{b/h_2}(N+1,\mu_F^2)
 \nn \\[2mm]
&\times& \left[ \;
\tilde{\omega}^{{\rm (res)}}_{ab}(N)
- \left. \tilde{\omega}^{{\rm (res)}}_{ab}(N)
\right|_{{\cal O}(\as^2)} \, \right] 
+\f{d\sigma^{\rm (NLO)}}{dQ^2dq_T^2} \;,
\eeeq
where $\tilde{\omega}^{{\rm (res)}}_{ab}$ is the resummed cross section for the partonic channel $ab\to \gamma^* c$ as given in Eq.~(\ref{eq:res}). In this way, NLO is fully taken into account, and the soft-gluon contributions beyond NLO are resummed to NLL. The procedure avoids
any double-counting of perturbative orders. 

Since the resummation is achieved in Mellin-moment space, one needs an inverse Mellin transform in order to obtain a resummed cross section in $\yT$ space. This requires a prescription for dealing with the Landau poles at $\lambda=1/2$ and $\lambda=1$ in Eqs.~(\ref{h1fun})-(\ref{fnll}) arising from the singularity in the perturbative strong coupling constant at scale $\Lambda_{\mathrm{QCD}}$. We employ the ``Minimal Prescription'' developed in Ref.~\cite{Catani:1996yz}, for which one uses the NLL expanded forms Eq.~(\ref{lndeltams}) and Eqs.(\ref{h1fun})-(\ref{fnll}), and 
chooses a Mellin contour in complex-$N$ space that lies to the left of the poles at $\lambda=1/2$ and $\lambda=1$ in the Mellin integrand.

\subsection*{Numerical results}

\begin{figure}
\begin{centering}
\includegraphics[scale=0.35]{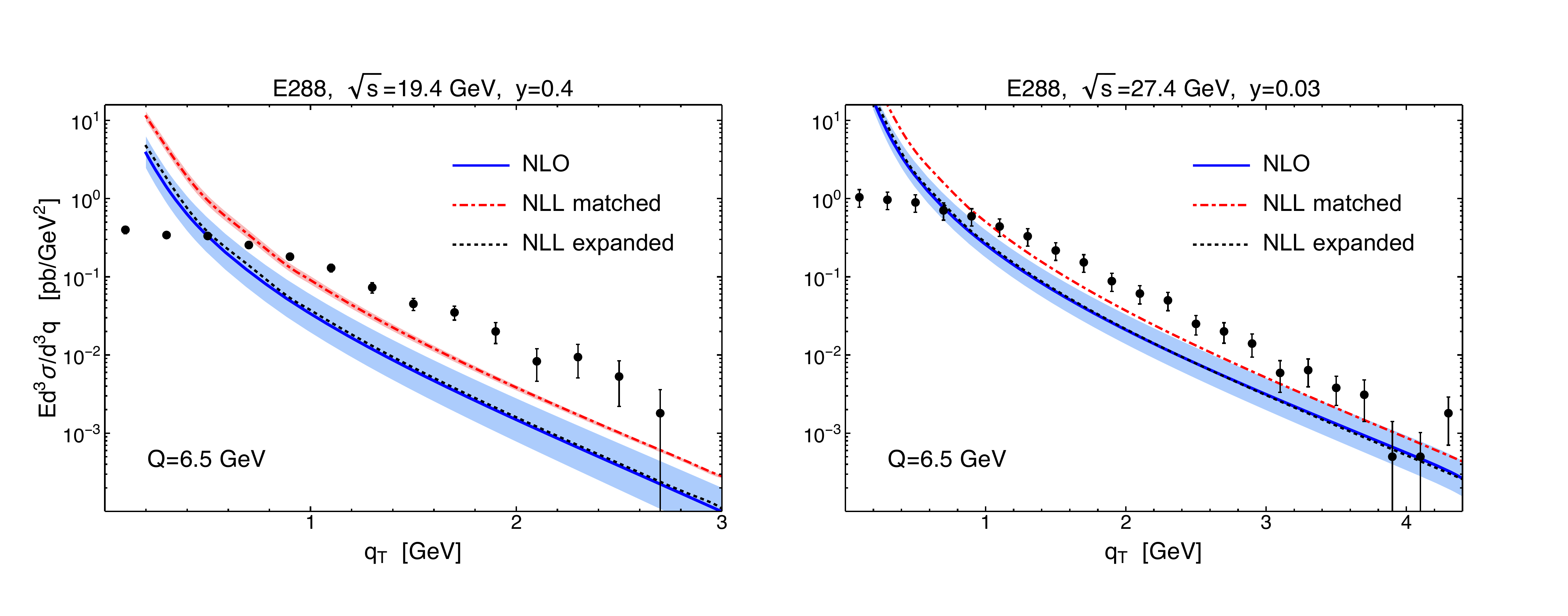}
\end{centering}
\caption{\label{E288thr} E288: experimental data vs. threshold-resummed predictions at NLL+NLO QCD for two different rapidity bins and two different center-of-mass energies.}
\end{figure}

\begin{figure}
\begin{centering}
\includegraphics[scale=0.35]{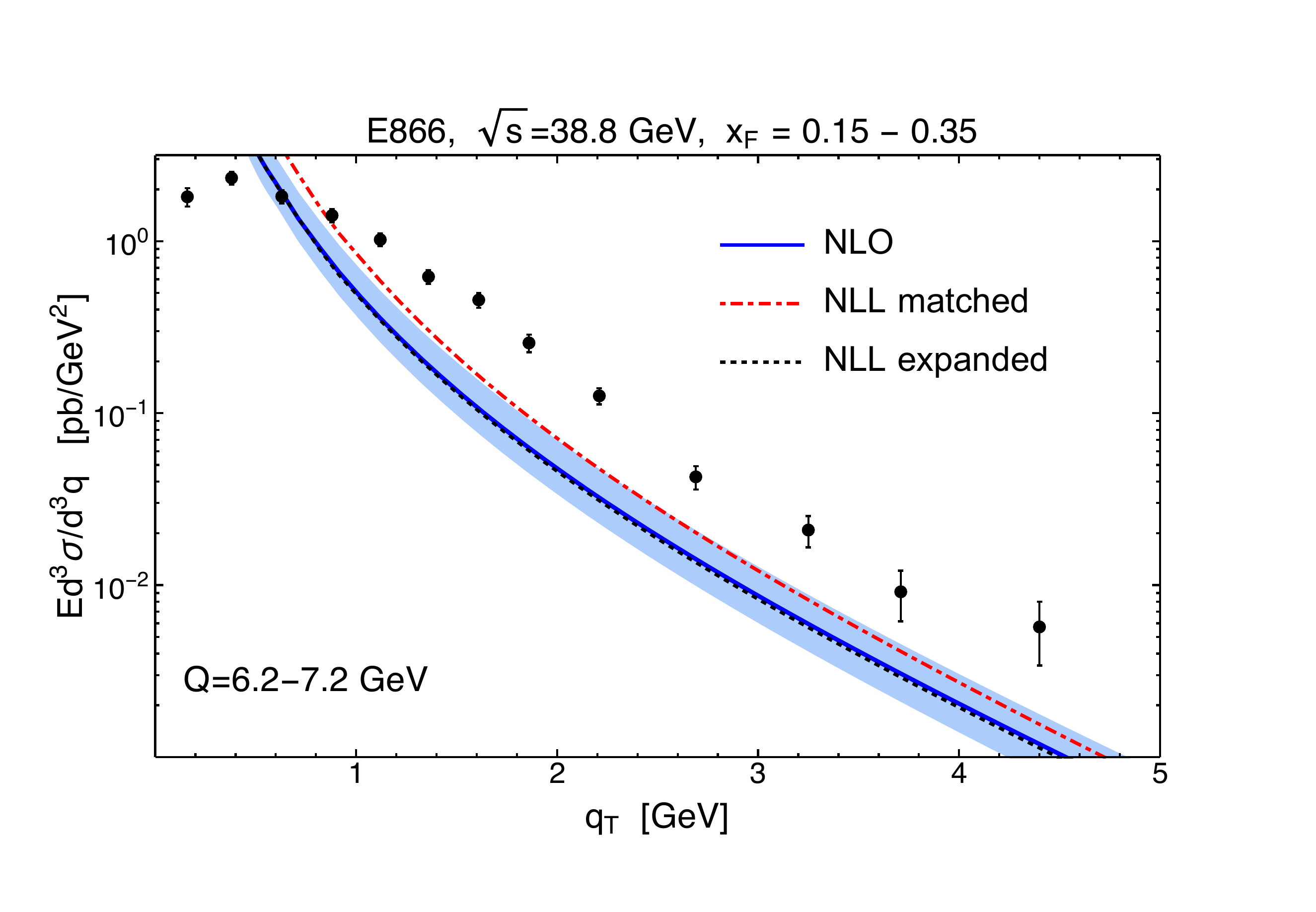}
\end{centering}
\caption{\label{E866thr} E866: experimental data vs. threshold-resummed predictions at NLL+NLO QCD for a selected ($x_{F},Q$) bin.}
\end{figure}

Numerical results for the above formalism are shown in Figs.~\ref{E288thr} and ~\ref{E866thr}, for a fixed value of Q and several values of  $\sqrt{s} $. We have chosen $\mu_F=\mu_R=Q$. To obtain predictions for a given experimental rapidity bin, we rescale the resummed cross section (which above was determined after integration over all $y$) by the ratio of NLO cross sections integrated over the $y$-bin used in experiment and integrated over all $y$, respectively:
\be
\int_{\Delta y_{{\mathrm{bin}}}} dy\,\f{ d\sigma^{\mathrm{res}}}{dQ^2dq_T^2dy}\,\approx\,\f{ d\sigma^{\mathrm{res}}}{dQ^2dq_T^2}\,\times\,
\frac{\int_{\Delta y_{{\mathrm{bin}}}} dy\,d\sigma^{\mathrm{NLO}}/dQ^2dq_T^2dy}{\int_{{\mathrm{all}\, y}} dy\,
d\sigma^{\mathrm{NLO}}/dQ^2dq_T^2dy}\,,
\ee
This approximation assumes that the rapidity dependence is similar at NLO and in the resummed case, an expectation that was confirmed in Ref.~\cite{Sterman:2000pt} for the closely related prompt-photon cross section.

We first notice from Figs.~\ref{E288thr} and ~\ref{E866thr} that the NLO expansion of the resummed formula (black dashed curve) accurately reproduces the NLO result (blue solid curve, with uncertainty bands). This provides some confidence that threshold resummation correctly describes the dominant parts of the cross section to all orders, and that subleading contributions not addressed by resummation are reasonably small. In the left part of Fig.~\ref{E288thr} we also show the scale uncertainty band for the NLL matched result (red dot-dashed curve), which is barely broad enough to be visible. Evidently, resummation leads to a strong reduction in scale dependence, as one would expect from a result that incorporates the dominant contributions to the cross section at all orders. 

Overall, we find a further significant increase of the cross section due to NLL resummation, with respect to the NLO results shown in Sec.~\ref{numerical}. The enhancement is more pronounced for the case of E288 than for E866 since, for a given $Q$, at E288 energy one is closer to threshold because of the lower c.m.s. energy. However, despite the increase, the NLL result unfortunately still remains well below the E288 and E866 experimental data at high $q_T$. We thus conclude that NLL high-$q_T$ threshold resummation is not able to lead to a satisfactory agreement with the data.


\section{Intrinsic-$k_{T}$ smearing and power corrections}

The factorized cross section given in Eq.~(\ref{eq:1}) receives corrections that are suppressed by inverse powers of $Q\sim q_T$. Little is known so far about the structure and size of such power corrections for the high-$q_T$ Drell--Yan cross section. It is an interesting question whether the discrepancies between perturbative predictions and the high-$q_T$ experimental data seen above might be explained by power corrections. We will try here to address this question from a phenomenological point of view. 

As a simple way of modeling power corrections we estimate below the impact of a non-perturbative partonic ``intrinsic'' transverse momentum $k_{T}$ on the Drell--Yan $q_{T}$ spectrum. Such an ``intrinsic-$k_{T}$ smearing'' is a phenomenological model that has been invoked in the early literature in cases where collinear factorization was found to underestimate transverse momentum spectra, like for inclusive prompt photon and pion production in hadronic collisions (see for instance \cite{Huston:1995vb,DAlesio:2004eso,Apanasevich:1998ki}). For inclusive processes such as these and the high-$q_T$ Drell--Yan process considered here, no general factorization theorem is known that would extend to arbitrary kinematics of the partonic process. For prompt photons, factorization has been established, however, for near-threshold kinematics and low $k_{T}$ in the framework of the ``joint resummation'' formalism~\cite{Laenen:2000ij,Laenen:2000de,Li:1998is}, and for high-energy (small-$x$) dynamics~\cite{Kimber:1999xc}. A technical challenge for all these approaches is the potential for an artificial singularity when the total transverse momentum of the initial state partons is comparable to the observed transverse momentum. A method for dealing with this issue was proposed in Ref.~\cite{Sterman:2004yk} and found to give rise to power corrections to the cross section. A full treatment of the Drell--Yan cross section may require implementation of perturbative joint resummation along with a study of corrections in inverse powers of $Q$ or $q_T$. Rather than pursuing this elaborate framework, for the purpose of obtaining a simple estimate of the potential size of such higher-order perturbative and power-suppressed non-perturbative effects, we resort to an implementation of a simple model of intrinsic-$k_{T}$ smearing that will be described now. 

\subsection*{Overview of the formalism}
\label{intrinsickt}

The collinear factorization formula for the process $h_1 h_2\to \gamma^* X$ may be adapted from Eq.~(\ref{eq:1}) and reads at LO $(\mathcal{O}\left(\alpha_{s}\right))$:
\begin{equation} 
\begin{split} 
E\frac{d^{3}\sigma}{d^{3}\mathbf{q}}\equiv\frac{d\sigma}{dy\,d^{2}\mathbf{q_{T}}} & = \sum_{a,b}\int dx_{a}\,dx_{b}\,f_{a/h_1}\left(x_{a},Q^{2}\right)f_{b/h_2}\left(x_{b},Q^{2}\right)\,  
\frac{d\hat{\sigma}^{ab\rightarrow\gamma^{*}c}}{d\hat{t}}\frac{\hat{s}}{\pi}\,\delta\left(\hat{s}+\hat{t}+\hat{u}-Q^{2}\right),
\label{collFac}
\end{split} 
\end{equation}
where as before the $f_{a/h}(x_{a},Q^{2})$ are the usual collinear PDFs for partons $a=q,\bar{q},g$ in hadron $h$. If one allows the incoming partons to have a small transverse momentum $\mathbf{k_{\mathit{T}}},$ Eq.~(\ref{collFac}) becomes \cite{DAlesio:2004eso}:
\begin{equation}
\begin{split} 
E\frac{d^{3}\sigma}{d^{3}\mathbf{q}} & =  \sum_{a,b}\int dx_{a}\,d^{2}\mathbf{k_{\mathit{\mathit{a}T}}}\,dx_{b}\,d^{2}\mathbf{k_{\mathit{b}\mathit{T}}}\,F_{a/h_1}\left(x_{a},\mathbf{k_{\mathit{aT}}},Q^{2}\right)F_{b/h_2}\left(x_{b},\mathbf{k_{\mathit{bT}}},Q^{2}\right)\\
 & \quad \times  \frac{\hat{s}}{x_{a}x_{b}s}\frac{d\hat{\sigma}^{ab\rightarrow\gamma^{*}c}}{d\hat{t}}\frac{\hat{s}}{\pi}\delta\left(\hat{s}+\hat{t}+\hat{u}-Q^{2}\right),
\label{kTsmear}
\end{split} 
\end{equation}
where the functions $F_{a/h}$ are a generalization of the PDFs, including a dependence on transverse momentum. Notice that the partonic Mandelstam invariants must be modified with the inclusion of $\mathbf{k_{\mathit{T}}},$ and consequently a factor $\hat{s}/(x_{a}x_{b}s)$ must be inserted to account for the modification of the partonic flux (see Appendix A of \cite{DAlesio:2004eso}). The modification of the partonic four-momenta is most often done according to two criteria: $\left(1\right)$ the partons remain on-shell: $p_{a\mu}p_{a}^{\mu}=0,$ and $\left(2\right)$ the light-cone momentum fractions retain the usual meaning, e.g.: $x_{a}=p_{a}^{+}/P_{a}^{+}.$ This leads to the following choice, in terms of Minkowski components \cite{Owens:1986mp,DAlesio:2004eso}: 
\begin{equation}
p_{a}^{\mu}\doteqdot\left(x_{a}\frac{\sqrt{s}}{2}+\frac{k_{aT}^{2}}{2x_{a}\sqrt{s}},\;\mathbf{\mathit{\mathbf{k_{\mathit{aT}}}}},\;x_{a}\frac{\sqrt{s}}{2}-\frac{k_{aT}^{2}}{2x_{a}\sqrt{s}}\right)\,,\label{parton4mom}
\end{equation}
and likewise for the other parton's momentum. Note that we use LO cross sections in Eq.~(\ref{kTsmear}) since a higher-order formulation is not really warranted for our simple model.

As mentioned above, the framework must become unreliable when $k_{aT}$ or $k_{bT}$ become of the order of the observed transverse momentum, and arguably well before. Large values of $k_{aT}$ can make the partonic Mandelstam in the denominators of the LO hard-scattering cross sections unphysically small. In \cite{DAlesio:2004eso}, the following condition was chosen to limit the size of, for example, $k_{aT}$:
\begin{equation}
k_{aT}<{\rm
  min}\left[x_{a}\sqrt{s},\sqrt{x_{a}\left(1-x_{a}\right)s}\right].\label{kTmax} 
\end{equation}
This ensures that each parton moves predominantly along the direction of its parent hadron, and that its energy does not exceed the hadron's energy. However, for $\sqrt{s}\simeq$ 40 GeV (E866 and E605 experiments), this condition implies that $k_{aT}$ may still reach values as high as 20 GeV. In our numerical analysis we therefore prefer to introduce an additional cutoff $k_{T{\mathrm{max}}}$  on both $k_{aT}$ and $k_{bT}$ and will test the dependence of the results on this cutoff.

\begin{figure}
\centering
\includegraphics[scale=0.4]{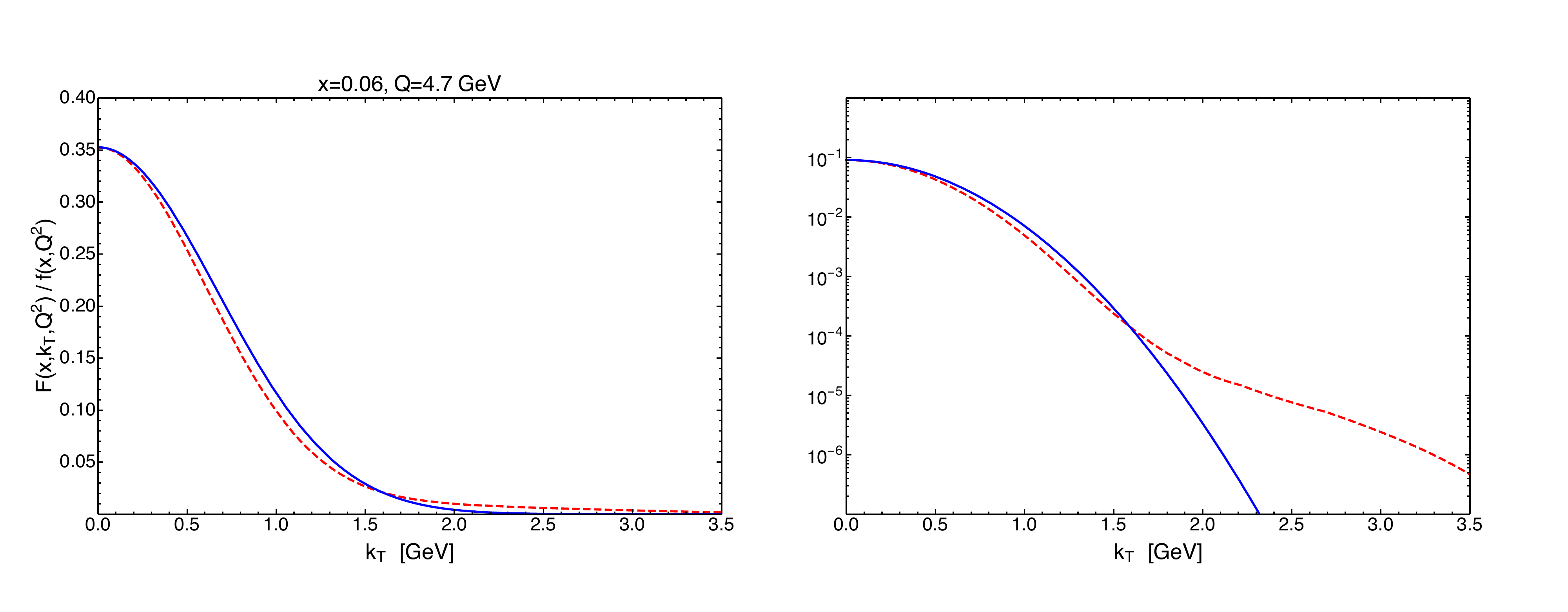}
\caption{\label{TMDcomparison}Comparison between the TMD of Ref.~\cite{Bacchetta:2017gcc}, evolved to the scale $Q$= 4.7 GeV and divided by its integral over $d^{2}\mathbf{k}_{T}$ (dashed line), with the Gaussian (\ref{gauss}) with $\left\langle k_{T}^{2}\right\rangle $ = (0.95 GeV)$^{2}$ (full line). Left panel: linear scale; Right panel: logarithmic scale.}
\end{figure}

For the generalized PDFs in Eq.~(\ref{kTsmear}), the most common choice is 
\begin{equation}
F_{a/h}\left(x_{a},\mathbf{k_{\mathit{aT}}},Q^{2}\right)=f_{a/h}\left(x_{a},Q^{2}\right)\,\frac{1}{\pi\left\langle k_{T}^{2}\right\rangle }\,
\exp\left[-\frac{k_{{a}T}^{2}}{\left\langle k_{T}^{2}\right\rangle}\right],\label{gauss}
\end{equation}
where $\left\langle k_{T}^{2}\right\rangle $ is independent of flavor\footnote{We remark that the initial parton ``$a$'' can also be a gluon. Every $k_{T}$-smearing model has to make an assumption for the average gluon transverse momentum, which is usually taken to be the same as that for the quarks. We note that perturbative resummations predict dependence of $\left\langle k_{T}^{2}\right\rangle$ on parton flavor~\cite{Sterman:2004yk}.} and momentum fraction $x_{a},$ but does depend logarithmically on $Q^{2}$ because of soft gluon radiation. Instead of Eq.~(\ref{gauss}), one could also consider using the transverse momentum dependent PDFs extracted from the low-q$_{T}$ spectra of Drell--Yan experiments (as given for instance in Refs.~\cite{Landry:2002ix,Konychev:2005iy,Bacchetta:2017gcc,Scimemi:2017etj}). However, these functions show a non-negligible tail at large $k_{T}$, where they lose physical meaning. Hence, if they are used inside a convolution such as Eq.~(\ref{kTsmear}), the result will strongly depend on the choice of the cutoff $k_{T{\mathrm{max}}}$, since the integrations~(\ref{kTsmear}) include contributions from this tail. This dependence will be mostly unphysical and is, in fact, precisely a manifestation of the artificial singularity arising in the partonic scattering functions at really large $k_{aT}$ and $k_{bT}$. For this reason, we stick with Eq.~(\ref{gauss}); however, we tune $\left\langle k_{T}^{2}\right\rangle$ to the width of the TMD PDFs taken from \cite{Bacchetta:2017gcc}, evolved to the given $Q^{2}$. This is shown in Fig.~\ref{TMDcomparison} where the dashed lines show the evolved TMD of Ref.~\cite{Bacchetta:2017gcc}, evolved to $Q$= 4.7 GeV, normalized by dividing by its integral over $d^{2}\mathbf{k}_{T}$. We compare it to a pure Gaussian with a width tuned in such a way that the two distributions become very similar, except for the high-$k_T$  tail. This ``equivalent Gaussian'' turns out to have a width of $\left\langle k_{T}^{2}\right\rangle $ = (0.95 GeV)$^{2}$. It is this Gaussian that we use for our numerical studies presented below.

Our choice of an $x$-{\it in}dependent Gaussian width in Eq.~(\ref{gauss}) is motivated by the fact that the $x$-dependence of $\left\langle k_{T}^{2}\right\rangle$ is still not well constrained in the present TMD fits~\cite{Bacchetta:2017gcc}. Different parametrizations have been proposed in the literature~\cite{Landry:2002ix,Konychev:2005iy}, including also $x$-independent choices~\cite{Anselmino:2013lza,DAlesio:2014mrz,Scimemi:2017etj}. A dependence of $\left\langle k_{T}^{2}\right\rangle$ on $x$ is a natural feature in the joint resummation formalism~\cite{Sterman:2004yk}. In any case, for the mostly exploratory study presented here, an $x$-independent value of $\left\langle k_{T}^{2}\right\rangle$ appears adequate. Since our goal is to give an upper limit for the $k_{T}$-smearing effects, we use the largest value of $\left\langle k_{T}^{2}\right\rangle$ found in~\cite{Bacchetta:2017gcc} (see Fig.~10 there), which occurs at $x=0.06$.

\subsection*{Numerical results}
\label{intrinsicnumerical}

\begin{figure}
\centering
\includegraphics[scale=0.4]{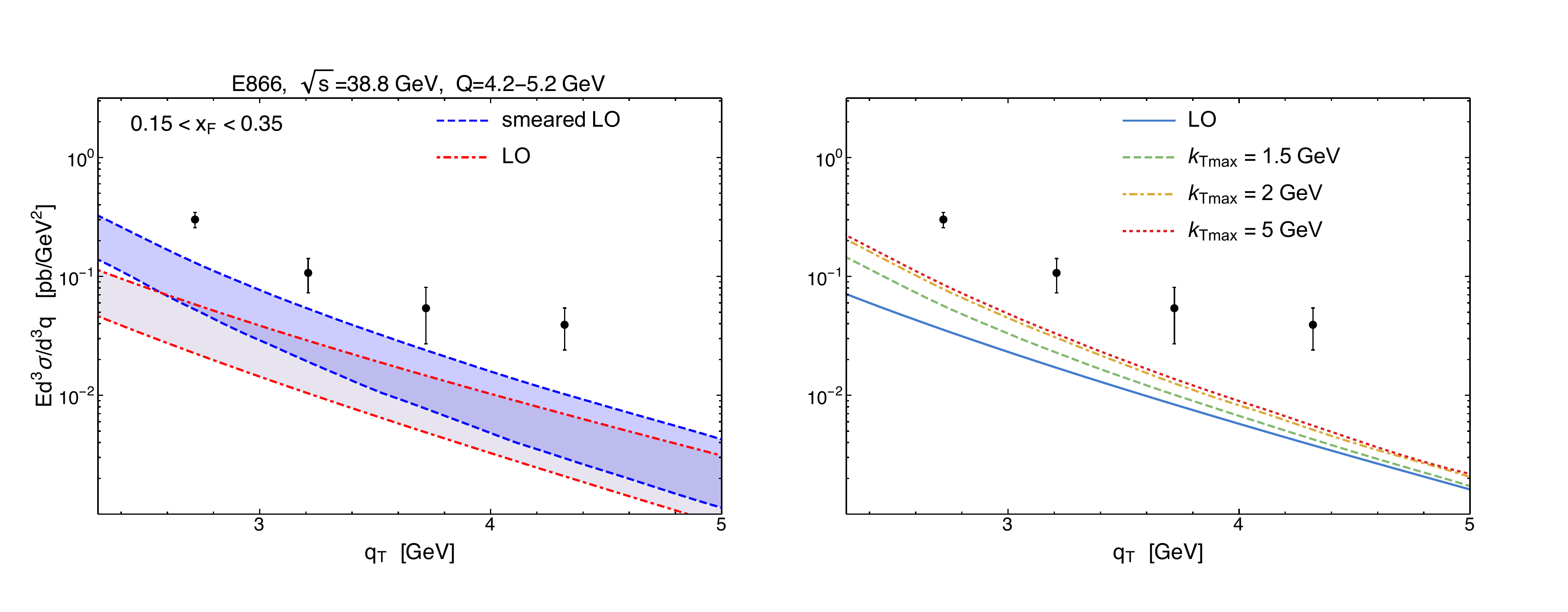}
\caption{\label{The-effect-of}Left panel: the effect of $k_{T}$-smearing (dashed blue lines), with the cutoff $k_{T{\mathrm{max}}}$ in Eq.~(\ref{kTsmear}) set to 2 GeV. The bands correspond to variation of factorization and renormalization scales between $Q/2$ and $2Q$. For comparison, the calculation in ordinary collinear factorization at LO is also shown (red dotted lines). Right panel: the effect of varying the cutoff $k_{T{\mathrm{max}}}$ in Eq.~(\ref{kTsmear}). Here the curves correspond to the central values $\mu_{R}=\mu_{F}=Q$. For $k_{T{\mathrm{max}}} \geq $ 2 GeV, which corresponds to the 99\% percentile of the gaussian in Eq.~(\ref{gauss}), independence from the cutoff is reached.} 
\end{figure} 

In Fig.~\ref{The-effect-of} we show the effect of $k_{T}$-smearing, Eq.~(\ref{kTsmear}), for E866 kinematics. The $\left\langle k_{T}^{2}\right\rangle $ of the Gaussian is taken as in Fig.~\ref{TMDcomparison}. The impact of smearing on the cross section overall remains mild, as long as the cutoff $k_{T{\mathrm{max}}}$ is chosen below 2~GeV. Especially the regime $q_{T}\simeq Q$ is only little affected by $k_{T}$-smearing. We conclude that, although $k_{T}$-smearing does somewhat improve the comparison with the data, its effects do not appear to be sufficiently large to lead to a satisfactory agreement. We note that at lower c.m.s. energies as relevant for E288, one is forced to choose smaller cutoffs since the reach in $q_{T}$ is more limited in these cases. 

\section{Conclusions}

We have shown that theoretical predictions based on fixed-order perturbation theory fail to describe Drell--Yan data from Fermilab and CERN ISR at large values $q_{T}\sim Q$ of the transverse momentum of the lepton pair, the experimental cross sections being
significantly larger than the theoretical ones. This is the region where collinear-factorized perturbation theory is expected to 
accurately describe the cross section. This disagreement is observed for several experiments, and across a range of different 
kinematics in  $x_{F}$, $y$ and $Q$, although admittedly the experimental uncertainties are in some cases quite large.  

We have on the other hand found an essentially satisfactory agreement between perturbative calculations and experimental points in the case of PHENIX data taken at $\sqrt{s}$ = 200 GeV, suggesting that the disagreement is present only in the fixed-target regime. 
Indeed, at yet higher energies, ATLAS Drell--Yan data ($\sqrt{s}$ = 8 TeV) have been shown to be consistently described by NNLO QCD supplemented with NNNLL resummation (see, for instance, Fig. 10 and Fig. 11 in~\cite{Bizon:2018foh}), even though some tension is still present in the lowest invariant mass bins (see Fig. 18 in~\cite{Aad:2015auj}).

Barring the possibility of sizable normalization uncertainties in the experiments,
it is important to identify the theoretical origins of the discrepancies observed in the fixed-target regime. We have first implemented perturbative threshold resummation and found that it improves the situation somewhat; a significant discrepancy remains, however. This leaves the investigation of power-suppressed corrections, which we have modeled by implementing a simple Gaussian intrinsic-$k_{T}$ smearing into the LO cross section. We find that this again helps somewhat, but does not lead to a satisfactory description of the data. 
Ultimately, a more detailed study of power corrections may be required in this case. 
Generically, on the basis of resummed perturbation theory~\cite{Sterman:2004yk}, one would expect {\it even} power corrections of the form $\lambda^2/\big(Q^2(1-y_T^2)^2\big)$, possibly modified by logarithms, where $\lambda$ is a hadronic mass scale. Given the kinematics of the experiments, it is hard to see how such corrections could become of the size needed for an adequate description of the data. 

Our findings are in line with those reported for the SIDIS cross section in Ref.~\cite{Gonzalez-Hernandez:2018ipj}. We close by stressing the importance of obtaining a thorough understanding of the full Drell--Yan and SIDIS $q_T$-spectra in the fixed-target regime. 
Low-$q_T$ Drell--Yan and SIDIS cross sections measured at fixed-target experiments are a prime source of information on TMDs. 
At present, the theoretical description for the important ``matching regime'' around $q_T=2$~GeV is not robust, as
we have argued. Given the shape of the experimental spectra, it appears that TMD physics may extend 
to such large $q_T$ and may well remain an important ingredient even beyond. This view is corroborated by the fact
that the $q_T$-integrated Drell--Yan cross section is well described by fixed-order perturbation theory at these energies. 
In any case, a reliable interpretation of data in terms of TMDs, including the matching to collinear physics, is only possible if the cross sections are theoretically understood over the full transverse-momentum range, which includes the regime of $q_T\sim Q$ we have addressed here.

\begin{acknowledgments}
AB, GB and FP acknowledge support from the European Research Council (ERC) under the European Union's Horizon 2020 research and innovation program (grant agreement No. 647981, 3DSPIN). This work has been supported in part by the Bundesministerium f\"{u}r Bildung und 
Forschung (BMBF) under grant no. 05P15VTCA1. We are grateful to the Institute
of Nuclear Physics for hospitality and support during program INT-18-3, where part of
this work has been carried out.

\end{acknowledgments}

\begin{appendix}

\section{LO cross sections}
\label{app:lo}

The explicit expressions for the Mellin moments of the LO partonic cross sections are given by
\ba
\tilde{\omega}^{(0)}_{q\bar{q}}(N) &=&C_F \, \frac{1-r}{1+r}\,
\left[  B\left(\frac{1}{2},N+1\right) {}_2F_1\left(\frac{1}{2},N+1,N+\frac{3}{2},\left(
 \frac{1-r}{1+r}\right)^2\right) \right. \nn \\[2mm] 
&-& \left. \frac{2r^2}{(1+r)^2} \,
 B\left(\frac{1}{2},N+2\right) {}_2F_1\left(\frac{1}{2},N+2,N+\frac{5}{2},\left(
 \frac{1-r}{1+r}\right)^2\right)
 \right. \nn \\[2mm]
&+& \left. \left(\frac{1-r}{1+r}\right)^2\, B\left(\frac{1}{2},N+3\right) {}_2F_1\left(\frac{1}{2},N+3,N+\frac{7}{2},\left(
 \frac{1-r}{1+r}\right)^2\right)\right] \, ,
\nn \\[2mm]
\tilde{\omega}^{(0)}_{qg}(N) &=& 
T_R\, \frac{1-r}{1+r}\,
\left[  B\left(\frac{1}{2},N+1\right) {}_2F_1\left(\frac{1}{2},N+1,N+\frac{3}{2},\left(
 \frac{1-r}{1+r}\right)^2\right) \right. \nn \\[2mm] 
&-& \left. \frac{3-4r^2}{(1+r)^2} \,
 B\left(\frac{1}{2},N+2\right) {}_2F_1\left(\frac{1}{2},N+2,N+\frac{5}{2},\left(
 \frac{1-r}{1+r}\right)^2\right)
 \right. \nn \\[2mm]
&+& \left. \frac{(4-r^2)(1-r)}{(1+r)^3}\, B\left(\frac{1}{2},N+3\right) {}_2F_1\left(\frac{1}{2},N+3,N+\frac{7}{2},\left(
 \frac{1-r}{1+r}\right)^2\right) \right. \nn \\[2mm]
&-& \left. 2\,\left(\frac{1-r}{1+r}\right)^3\, B\left(\frac{1}{2},N+4\right) {}_2F_1\left(\frac{1}{2},N+4,N+\frac{9}{2},\left(
 \frac{1-r}{1+r}\right)^2\right)\right] \, ,
\ea
where $r=q_T/\sqrt{q_T^2+Q^2}$, $C_F=4/3$, $T_R=1/2$, and where ${}_2 F_1$ is the hypergeometric function.

\section{One loop coefficients }
\label{app:coeff}

The one-loop coefficients $C_{ab \to \gamma^* c}^{(1)}$ for the subprocesses read
\ba
C_{q\bar{q} \to \gamma^* g}^{(1)}&=& 
 \left(C_F \left(r^2+2 r+2\right)-\frac{1}{2}
    C_A r (r+2)\right)\,\text{Li}_2(1-r)+\frac{1}{2} \left(r^2+2 r+2\right) (C_A-2 C_F)
 \,   \text{Li}_2\left(\frac{1-r}{1+r}\right)\nn\\[2mm]
    &+&\pi b_0
    \ln \left(\frac{\mu_R^2}{Q_0^2}\right)-\frac{3}{2} C_F \ln
    \left(\frac{\mu_F^2}{Q_0^2}\right)+\frac{1}{36} \left(C_A \left(9 r^2-9 r+67\right)+2
    \left(9 C_F \left(r^3-8\right)-5N_f\right)\right)\nn\\[2mm]
    &+&\frac{1}{4} \left(r^2+2
    r+2\right) (C_A-2 C_F) \ln ^2(1+r)\nn\\[2mm]
    &-&\frac{1}{8} (C_A-2 C_F)
    \left(r^2 (\ln (16)-3)+r (\ln (256)-2)+1+\ln (256)\right) \ln (1+r)\nn\\[2mm]
    &-&\frac{1}{2}
    \left(r^2+2 r+2\right) (C_A-2 C_F) \ln (1-r) \ln(1+r)+\frac{1}{2} \ln
    r \left(-C_A r^3-C_F \left(r^4-2 r^3-4 r^2+3\right)\right)\nn\\[2mm]
    &+&\frac{1}{8}
    \ln (1-r) \left(C_A \left(4 r^3+r^2 (\ln (16)-3)+r (\ln (256)-2)+1+\ln
    (256)\right)\right.\nn\\[2mm]
    &+&\left.2 C_F \left(2 r^4-4 r^3-r^2 (5+\ln (16))+r (2-8 \ln (2))+5-8 \ln
    (2)\right)\right)\nn\\[2mm]
    &+&\pi ^2 \left(\frac{5
    C_F}{6}-\frac{C_A}{4}\right)-\frac{1}{2} C_A \ln ^2(r)+C_A
    \ln (1-r) \ln (r)\,,
\ea
and
\ba
C_{qg \to \gamma^*q}^{(1)} &=& -\frac{ 3 C_A r^2+C_A-2 C_F
    \left(7 r^2+2\right)}{2(4 r^2+1)}\,\text{Li}_2(1-r)+
    \frac{C_A \left(r^2-1\right)-10 C_F r^2}{2(4 r^2+1)}\,
    \text{Li}_2\left(\frac{1-r}{1+r}\right)\nn\\[2mm]
    &+&\pi b_0
    \ln \left(\frac{\mu_R^2}{Q_0^2}\right)-\left(
    \pi b_0 + \frac{3}{4}C_F\right) \ln
    \left(\frac{\mu_F^2}{Q_0^2}\right)+
    \frac{6 C_A \left(-2
    r^3+r+1\right)+3 C_F \left(10 r^3-31 r^2-3 r-11\right)}{12(4 r^2+1)}\nn\\[2mm]
    &+&\frac{\pi ^2
    \left(17 C_A r^2+2 C_A+2 C_F r^2+5 C_F\right)}{12(4 r^2+1)}+\frac{\left(r^2+1\right) (C_A-2 C_F) \ln ^2 r}{2(4 r^2+1)}\nn\\[2mm]
    &+&\frac{\ln ^2(1+r) \left(C_A \left(r^2-1\right)-10 C_F
    r^2\right)}{4(4 r^2+1)}-\frac{\left(r^2+1\right) (C_A-2 C_F) \ln (1-r) \ln
    r}{4 r^2+1}\nn\\[2mm]
    &+&\frac{\ln (r+1) \left(C_A \left(r^2 (2-4 \ln (2))+4 r+2+\ln
    (16)\right)+C_F \left(5 r^2 (\ln (256)-1)-2 r+3\right)\right)}{8(4 r^2+1)}\nn\\[2mm]
    &+&\frac{\ln (1-r) \ln (1+r) \left(-C_A r^2+C_A+10 C_F
    r^2\right)}{2(4 r^2+1)}+\frac{\ln r \left(C_A \left(2 r^2-r-3\right)
    r^2+C_F \left(-5 r^4+8 r^3+r^2-3\right)\right)}{2(4 r^2+1)}\nn\\[2mm]
    &+&\frac{\ln (1-r)
    \left(C_F \left(20 r^4-32 r^3+r^2 (1-40 \ln (2))+2 r+9\right)-2 C_A
    \left(r^2-1\right) \left(4 r^2-2 r-1-\ln (4)\right)\right)}{8(4 r^2+1)}\nn\\
\ea
where $Q_0^2=q_T(q_T+m_T)$ and $\b0=(11 C_A - 2 N_f )/12 \pi$. In the limit $Q\to 0$ (or $r\to 1$) these coefficients agree with the ones found for prompt-photon production in Ref.~\cite{Catani:1998tm}. 

\section{LL and NLL functions}
\label{app:hf}

The explicit expressions for the LL and NLL functions in Eq.~(\ref{lndeltams}) are:
\ba
\label{h1fun}
h_a^{(1)}(\la)& =&\frac{A_a^{(1)}}{2\pi \b0 \la} 
\left[ 2 \la+(1-2 \la)\ln(1-2\la)\right] \;,\nn\\[2mm]
h_a^{(2)}(\la,Q_0^2/\mu^2_R;Q_0^2/\mu_{F}^2) 
&=&-\;\f{A_a^{(2)}}{2\pi^2 \b0^2 } \left[ 2 \la+\ln(1-2\la)\right] \,+\, \f{A_a^{(1)} b_1}{2\pi \b0^3} 
\left[2 \la+\ln(1-2\la)+\f{1}{2} \ln^2(1-2\la)\right]\nn \\[2mm]
\label{h2fun}
&&+\; \f{A_a^{(1)}}{2\pi \b0}\left[2 \la+\ln(1-2\la) \right]  
\ln\f{Q_0^2}{\mu^2_R}\,-\,\f{A_a^{(1)}}{\pi \b0} \,\la \ln\f{Q_0^2}{\mu^2_{F}} \;,  
\ea
and
\ba
\label{fll}
f_a^{(1)}(\lambda) &=&h_a^{(1)}(\la/2)-h_a^{(1)}(\la)\; ,\nn \\[2mm]
\label{fnll}
f_a^{(2)}(\lambda,Q_0^2/\mu^2_R) &=&2\,h_a^{(2)}(\la/2,Q_0^2/\mu^2_R,1)-h_a^{(2)}(\la,Q_0^2/\mu^2_R,1)+
\frac{B_a^{(1)}}{2\pi b_0}\ln(1-\lambda) \; ,
\ea
where
\ba
\b0&=& \frac{11 C_A - 2 N_f }{12 \pi}\,,\nn\\[2mm]
b_1&=&  \frac{1}{24 \pi^2} 
\left( 17 C_A^2 - 5 C_A N_f - 3 C_F N_f \right) \,,
\label{b1coef}
\ea
are the first two coefficients of the QCD $\beta$ function. 

\end{appendix}

\bibliographystyle{h-physrev5}
\bibliography{References}

\begin{thebibliography}{10}

\bibitem{Drell:1970wh}
S.~D. Drell and T.-M. Yan,
\newblock Phys. Rev. Lett. {\bf 25}, 316 (1970),
\newblock [Erratum: Phys. Rev. Lett.25,902(1970)].

\bibitem{Peng:2014hta}
J.-C. Peng and J.-W. Qiu,
\newblock Prog. Part. Nucl. Phys. {\bf 76}, 43 (2014), arXiv:1401.0934.

\bibitem{Collins:1989gx}
J.~C. Collins, D.~E. Soper, and G.~F. Sterman,
\newblock Adv. Ser. Direct. High Energy Phys. {\bf 5}, 1 (1989),
  arXiv:hep-ph/0409313.

\bibitem{Butterworth:2015oua}
J.~Butterworth {\em et~al.},
\newblock J. Phys. {\bf G43}, 023001 (2016), arXiv:1510.03865.

\bibitem{Accardi:2016ndt}
A.~Accardi {\em et~al.},
\newblock Eur. Phys. J. {\bf C76}, 471 (2016), arXiv:1603.08906.

\bibitem{Landry:1999an}
F.~Landry, R.~Brock, G.~Ladinsky, and C.~P. Yuan,
\newblock Phys. Rev. {\bf D63}, 013004 (2001), arXiv:hep-ph/9905391.

\bibitem{Landry:2002ix}
F.~Landry, R.~Brock, P.~M. Nadolsky, and C.~P. Yuan,
\newblock Phys. Rev. {\bf D67}, 073016 (2003), arXiv:hep-ph/0212159.

\bibitem{DAlesio:2004eso}
U.~D'Alesio and F.~Murgia,
\newblock Phys. Rev. {\bf D70}, 074009 (2004), arXiv:hep-ph/0408092.

\bibitem{Konychev:2005iy}
A.~V. Konychev and P.~M. Nadolsky,
\newblock Phys. Lett. {\bf B633}, 710 (2006), arXiv:hep-ph/0506225.

\bibitem{DAlesio:2014mrz}
U.~D'Alesio, M.~G. Echevarria, S.~Melis, and I.~Scimemi,
\newblock JHEP {\bf 11}, 098 (2014), arXiv:1407.3311.

\bibitem{Su:2014wpa}
P.~Sun, J.~Isaacson, C.~P. Yuan, and F.~Yuan,
\newblock Int. J. Mod. Phys. {\bf A33}, 1841006 (2018), arXiv:1406.3073.

\bibitem{Pasquini:2014ppa}
B.~Pasquini and P.~Schweitzer,
\newblock Phys. Rev. {\bf D90}, 014050 (2014), arXiv:1406.2056.

\bibitem{Bacchetta:2017gcc}
A.~Bacchetta, F.~Delcarro, C.~Pisano, M.~Radici, and A.~Signori,
\newblock JHEP {\bf 06}, 081 (2017), arXiv:1703.10157.

\bibitem{Scimemi:2017etj}
I.~Scimemi and A.~Vladimirov,
\newblock Eur. Phys. J. {\bf C78}, 89 (2018), arXiv:1706.01473.

\bibitem{Ceccopieri:2018nop}
F.~A. Ceccopieri, A.~Courtoy, S.~Noguera, and S.~Scopetta,
\newblock Eur. Phys. J. {\bf C78}, 644 (2018), arXiv:1801.07682.

\bibitem{Berger:2001wr}
E.~L. Berger, J.-w. Qiu, and X.-f. Zhang,
\newblock Phys. Rev. {\bf D65}, 034006 (2002), arXiv:hep-ph/0107309.

\bibitem{Ridder:2016nkl}
A.~Gehrmann-De~Ridder, T.~Gehrmann, E.~W.~N. Glover, A.~Huss, and T.~A. Morgan,
\newblock JHEP {\bf 07}, 133 (2016), arXiv:1605.04295.

\bibitem{Gonsalves:1989ar}
R.~J. Gonsalves, J.~Pawlowski, and C.-F. Wai,
\newblock Phys. Rev. {\bf D40}, 2245 (1989).

\bibitem{Mirkes:1992hu}
E.~Mirkes,
\newblock Nucl. Phys. {\bf B387}, 3 (1992).

\bibitem{Melnikov:2006kv}
K.~Melnikov and F.~Petriello,
\newblock Phys. Rev. {\bf D74}, 114017 (2006), arXiv:hep-ph/0609070.

\bibitem{Catani:2009sm}
S.~Catani, L.~Cieri, G.~Ferrera, D.~de~Florian, and M.~Grazzini,
\newblock Phys. Rev. Lett. {\bf 103}, 082001 (2009), arXiv:0903.2120.

\bibitem{Gavin:2010az}
R.~Gavin, Y.~Li, F.~Petriello, and S.~Quackenbush,
\newblock Comput. Phys. Commun. {\bf 182}, 2388 (2011), arXiv:1011.3540.

\bibitem{Bozzi:2010xn}
G.~Bozzi, S.~Catani, G.~Ferrera, D.~de~Florian, and M.~Grazzini,
\newblock Phys. Lett. {\bf B696}, 207 (2011), arXiv:1007.2351.

\bibitem{Bozzi:2008bb}
G.~Bozzi, S.~Catani, G.~Ferrera, D.~de~Florian, and M.~Grazzini,
\newblock Nucl. Phys. {\bf B815}, 174 (2009), arXiv:0812.2862.

\bibitem{Becher:2011xn}
T.~Becher, M.~Neubert, and D.~Wilhelm,
\newblock JHEP {\bf 02}, 124 (2012), arXiv:1109.6027.

\bibitem{Collins:1984kg}
J.~C. Collins, D.~E. Soper, and G.~F. Sterman,
\newblock Nucl. Phys. {\bf B250}, 199 (1985).

\bibitem{Boer:2006eq}
D.~Boer and W.~Vogelsang,
\newblock Phys. Rev. {\bf D74}, 014004 (2006), arXiv:hep-ph/0604177.

\bibitem{Berger:2007si}
E.~L. Berger, J.-W. Qiu, and R.~A. Rodriguez-Pedraza,
\newblock Phys. Lett. {\bf B656}, 74 (2007), arXiv:0707.3150.

\bibitem{Peng:2015spa}
J.-C. Peng, W.-C. Chang, R.~E. McClellan, and O.~Teryaev,
\newblock Phys. Lett. {\bf B758}, 384 (2016), arXiv:1511.08932.

\bibitem{Lambertsen:2016wgj}
M.~Lambertsen and W.~Vogelsang,
\newblock Phys. Rev. {\bf D93}, 114013 (2016), arXiv:1605.02625.

\bibitem{Gonzalez-Hernandez:2018ipj}
J.~O. Gonzalez-Hernandez, T.~C. Rogers, N.~Sato, and B.~Wang,
\newblock Phys. Rev. {\bf D98}, 114005 (2018), arXiv:1808.04396.

\bibitem{Arnold:1990yk}
P.~B. Arnold and R.~P. Kauffman,
\newblock Nuclear Physics B {\bf 349}, 381  (1991).

\bibitem{Boglione:2014oea}
M.~Boglione, J.~O. Gonzalez~Hernandez, S.~Melis, and A.~Prokudin,
\newblock JHEP {\bf 02}, 095 (2015), arXiv:1412.1383.

\bibitem{Collins:2016hqq}
J.~Collins {\em et~al.},
\newblock Phys. Rev. {\bf D94}, 034014 (2016), arXiv:1605.00671.

\bibitem{Echevarria:2018qyi}
M.~G. Echevarria, T.~Kasemets, J.-P. Lansberg, C.~Pisano, and A.~Signori,
\newblock Phys. Lett. {\bf B781}, 161 (2018), arXiv:1801.01480.

\bibitem{Bacchetta:2015ora}
A.~Bacchetta, M.~G. Echevarria, P.~J.~G. Mulders, M.~Radici, and A.~Signori,
\newblock JHEP {\bf 11}, 076 (2015), arXiv:1508.00402.

\bibitem{Ito:1980ev}
A.~S. Ito {\em et~al.},
\newblock Phys. Rev. {\bf D23}, 604 (1981).

\bibitem{Dulat:2015mca}
S.~Dulat {\em et~al.},
\newblock Phys. Rev. {\bf D93}, 033006 (2016), arXiv:1506.07443.

\bibitem{Hawker:1998ty}
NuSea, E.~A. Hawker {\em et~al.},
\newblock Phys. Rev. Lett. {\bf 80}, 3715 (1998), arXiv:hep-ex/9803011.

\bibitem{Webb:2003bj}
J.~C. Webb,
\newblock {\em {Measurement of continuum dimuon production in 800-GeV/C proton
  nucleon collisions}},
\newblock PhD thesis, New Mexico State U., 2003, arXiv:hep-ex/0301031.

\bibitem{Martin:2009iq}
A.~D. Martin, W.~J. Stirling, R.~S. Thorne, and G.~Watt,
\newblock Eur. Phys. J. {\bf C63}, 189 (2009), arXiv:0901.0002.

\bibitem{Ball:2017nwa}
NNPDF, R.~D. Ball {\em et~al.},
\newblock Eur. Phys. J. {\bf C77}, 663 (2017), arXiv:1706.00428.

\bibitem{Lai:2010vv}
H.-L. Lai {\em et~al.},
\newblock Phys. Rev. {\bf D82}, 074024 (2010), arXiv:1007.2241.

\bibitem{Ball:2012cx}
R.~D. Ball {\em et~al.},
\newblock Nucl. Phys. {\bf B867}, 244 (2013), arXiv:1207.1303.

\bibitem{Antreasyan:1980yb}
D.~Antreasyan {\em et~al.},
\newblock Phys. Rev. Lett. {\bf 45}, 863 (1980).

\bibitem{Antreasyan:1981uv}
D.~Antreasyan {\em et~al.},
\newblock Phys. Rev. Lett. {\bf 47}, 12 (1981).

\bibitem{Antreasyan:1981eg}
{D. Antreasyan et al.},
\newblock Phys. Rev. Lett. {\bf 48}, 302 (1982).

\bibitem{Gavin:1995ch}
S.~Gavin {\em et~al.},
\newblock Int. J. Mod. Phys. {\bf A10}, 2961 (1995), arXiv:hep-ph/9502372.

\bibitem{Szczurek:2008ga}
A.~Szczurek and G.~Slipek,
\newblock Phys. Rev. {\bf D78}, 114007 (2008), arXiv:0808.1360.

\bibitem{Kovarik:2015cma}
K.~Kovarik {\em et~al.},
\newblock Phys. Rev. {\bf D93}, 085037 (2016), arXiv:1509.00792.

\bibitem{Moreno:1990sf}
G.~Moreno {\em et~al.},
\newblock Phys. Rev. {\bf D43}, 2815 (1991).

\bibitem{Aidala:2018ajl}
PHENIX, C.~Aidala {\em et~al.},
\newblock Submitted to: Phys. Rev. D  (2018), arXiv:1805.02448.

\bibitem{deFlorian:2005yj}
D.~de~Florian and W.~Vogelsang,
\newblock Phys. Rev. {\bf D71}, 114004 (2005), arXiv:hep-ph/0501258.

\bibitem{Hinderer:2018nkb}
P.~Hinderer, F.~Ringer, G.~Sterman, and W.~Vogelsang,
\newblock (2018), arXiv:1812.00915.

\bibitem{Kidonakis:1999ur}
N.~Kidonakis and V.~Del~Duca,
\newblock Phys. Lett. {\bf B480}, 87 (2000), arXiv:hep-ph/9911460.

\bibitem{Kidonakis:2003xm}
N.~Kidonakis and A.~Sabio~Vera,
\newblock JHEP {\bf 02}, 027 (2004), arXiv:hep-ph/0311266.

\bibitem{Gonsalves:2005ng}
R.~J. Gonsalves, N.~Kidonakis, and A.~Sabio~Vera,
\newblock Phys. Rev. Lett. {\bf 95}, 222001 (2005), arXiv:hep-ph/0507317.

\bibitem{Kidonakis:2014zva}
N.~Kidonakis and R.~J. Gonsalves,
\newblock Phys. Rev. {\bf D89}, 094022 (2014), arXiv:1404.4302.

\bibitem{Muselli:2017bad}
C.~Muselli, S.~Forte, and G.~Ridolfi,
\newblock JHEP {\bf 03}, 106 (2017), arXiv:1701.01464.

\bibitem{deFlorian:2005fzc}
D.~de~Florian, A.~Kulesza, and W.~Vogelsang,
\newblock JHEP {\bf 02}, 047 (2006), arXiv:hep-ph/0511205.

\bibitem{Sterman:1986aj}
G.~F. Sterman,
\newblock Nucl. Phys. {\bf B281}, 310 (1987).

\bibitem{Catani:1990rp}
S.~Catani and L.~Trentadue,
\newblock Nucl. Phys. {\bf B353}, 183 (1991).

\bibitem{Catani:1998tm}
S.~Catani, M.~L. Mangano, and P.~Nason,
\newblock JHEP {\bf 07}, 024 (1998), arXiv:hep-ph/9806484.

\bibitem{Catani:1999hs}
S.~Catani, M.~L. Mangano, P.~Nason, C.~Oleari, and W.~Vogelsang,
\newblock JHEP {\bf 03}, 025 (1999), arXiv:hep-ph/9903436.

\bibitem{Kodaira:1982az}
J.~Kodaira and L.~Trentadue,
\newblock Phys. Lett. {\bf 123B}, 335 (1983).

\bibitem{Catani:1988vd}
S.~Catani, E.~D'Emilio, and L.~Trentadue,
\newblock Phys. Lett. {\bf B211}, 335 (1988).

\bibitem{Catani:1996yz}
S.~Catani, M.~L. Mangano, P.~Nason, and L.~Trentadue,
\newblock Nucl. Phys. {\bf B478}, 273 (1996), arXiv:hep-ph/9604351.

\bibitem{Sterman:2000pt}
G.~F. Sterman and W.~Vogelsang,
\newblock JHEP {\bf 02}, 016 (2001), arXiv:hep-ph/0011289.

\bibitem{Huston:1995vb}
J.~Huston {\em et~al.},
\newblock Phys. Rev. {\bf D51}, 6139 (1995), arXiv:hep-ph/9501230.

\bibitem{Apanasevich:1998ki}
L.~Apanasevich {\em et~al.},
\newblock Phys. Rev. {\bf D59}, 074007 (1999), arXiv:hep-ph/9808467.

\bibitem{Laenen:2000ij}
E.~Laenen, G.~F. Sterman, and W.~Vogelsang,
\newblock Phys. Rev. {\bf D63}, 114018 (2001), arXiv:hep-ph/0010080.

\bibitem{Laenen:2000de}
E.~Laenen, G.~F. Sterman, and W.~Vogelsang,
\newblock Phys. Rev. Lett. {\bf 84}, 4296 (2000), arXiv:hep-ph/0002078.

\bibitem{Li:1998is}
H.-n. Li,
\newblock Phys. Lett. {\bf B454}, 328 (1999), arXiv:hep-ph/9812363.

\bibitem{Kimber:1999xc}
M.~A. Kimber, A.~D. Martin, and M.~G. Ryskin,
\newblock Eur. Phys. J. {\bf C12}, 655 (2000), arXiv:hep-ph/9911379.

\bibitem{Sterman:2004yk}
G.~F. Sterman and W.~Vogelsang,
\newblock Phys. Rev. {\bf D71}, 014013 (2005), arXiv:hep-ph/0409234.

\bibitem{Owens:1986mp}
J.~F. Owens,
\newblock Rev. Mod. Phys. {\bf 59}, 465 (1987).

\bibitem{Anselmino:2013lza}
M.~Anselmino, M.~Boglione, J.~O. Gonzalez~Hernandez, S.~Melis, and A.~Prokudin,
\newblock JHEP {\bf 04}, 005 (2014), arXiv:1312.6261.

\bibitem{Bizon:2018foh}
W.~Bizon {\em et~al.},
\newblock (2018), arXiv:1805.05916.

\bibitem{Aad:2015auj}
ATLAS, G.~Aad {\em et~al.},
\newblock Eur. Phys. J. {\bf C76}, 291 (2016), arXiv:1512.02192.

\end{thebibliography}

\end{document}